\documentclass[11pt]{amsart}

\usepackage{amscd}
\usepackage{amsmath}
\usepackage{graphicx}
\usepackage{amsfonts}
\usepackage{amssymb}
\begin{document}

\title[Positive finite rank elementary operators]
{Positive finite rank elementary operators and characterizing
entanglement of states}

\author{Xiaofei Qi}
\address[Xiaofei Qi]{
Department of Mathematics, Shanxi University , Taiyuan 030006, P. R.
of China;} \email{qixf1980@126.com}

\author{Jinchuan Hou}
\address{Department of
Mathematics\\
Taiyuan University of Technology\\
 Taiyuan 030024,
  P. R. of China}
\email{jinchuanhou@yahoo.com.cn}


\thanks{This work is partially supported by National Natural Science
Foundation of China (No. 10771157), Research Grant to Returned
Scholars of Shanxi (2007-38) and Foundation of Shanxi University.}

\begin{abstract}

In this paper, a class of indecomposable positive finite rank
elementary operators of order $(n,n)$ are constructed. This allows
us to give a simple necessary and sufficient criterion for
separability of pure states in bipartite systems of any dimension in
terms of positive elementary operators of order $(2,2)$ and get some
new mixed entangled states that can not be detected by the positive
partial transpose (PPT) criterion and the realignment criterion.

\end{abstract}
\maketitle

\section{Introduction}

Entanglement is a basic physical resource to realize various quantum
information  and quantum communication tasks such as quantum
cryptography, teleportation, dense coding and key distribution
\cite{NC}. Let $H$ and $K$ be separable complex Hilbert spaces.
Recall that a quantum state is a density operator $\rho\in{\mathcal
B}(H\otimes K)$ which is positive and has trace 1. Denote by
${\mathcal S}(H)$ the set of all states on $H\otimes K$. If $H$ and
$K$ are finite dimensional, $\rho\in{\mathcal S}(H\otimes K)$ is
said to be separable if $\rho$ can be written as
$$\rho=\sum_{i=1}^k p_i \rho_i\otimes \sigma _i,$$
where $\rho_i$ and $\sigma_i$ are states on $H$ and $K$
respectively, and $p_i$ are positive numbers with $\sum
_{i=1}^kp_i=1$. Otherwise, $\rho$ is said to be inseparable or
entangled (ref. \cite{BZ, NC}). For the case that at least one of
$H$ and $K$ is of infinite dimension,  by  Werner \cite{W},  a state
$\rho$ acting on $H\otimes K$ is called separable if it can be
approximated in the trace norm by the states of the form
$$\sigma=\sum_{i=1}^n p_i \rho_i\otimes \sigma _i,$$
where $\rho_i$ and $\sigma_i$ are states on $H$ and $K$
respectively, and $p_i$ are positive numbers with
$\sum_{i=1}^np_i=1$. Otherwise, $\rho$ is called an entangled state.

It is very important but also difficult to determine whether or not
a state in a composite system is separable. For $2\times 2$ and
$2\times 3$ systems, that is, for the case $\dim H =\dim K = 2$ or
$\dim H = 2,\ \dim K = 3$, a state is separable if and only if it is
a positive partial transpose (PPT) state (see \cite{Hor, Pe}), but
it has no efficiency for PPT entangled states appearing in the
higher dimensional systems. In \cite{CW}, the realignment criterion
for separability in finite-dimensional systems was found. A most
general approach to characterize quantum entanglement is based on
the notion of entanglement witnesses (see \cite{Hor}). A Hermitian
operator $W$ acting on $H\otimes K$ is said to be an entanglement
witness (briefly, EW), if $W$ is not positive and ${\rm
Tr}(W\sigma)\geq 0$ holds for all separable states $\sigma$. Thus,
if $W$ is an EW, then there exists an entangled state $\rho$ such
that ${\rm Tr}(W\rho) < 0$ (that is, the entanglement of $\rho$ can
be detected by $W$). It was shown that, a state is entangled if and
only if it is detected by some entanglement witnesses \cite{Hor}.
However, constructing entanglement witnesses is a hard task. There
was a considerable effort in constructing and analyzing the
structure of entanglement witnesses for finite and infinite
dimensional systems \cite{B,TG,CK,JB,HQ} (see also \cite{HHH1} for a
review). Recently, Hou and Qi in \cite{HQ} showed that every
entangled state  in a bipartite system can be detected by some
entanglement witness $W$  of the form $W=cI+T$ with $I$ the identity
operator, $c$ a nonnegative number and $T$ a finite rank
self-adjoint operator.

 It is obvious that if $\rho$ is a
state on $H\otimes K$, then for every completely positive linear map
$\Phi :{\mathcal B}(H)\rightarrow {\mathcal B}(K)$, the operator
$(\Phi\otimes I)\rho\in{\mathcal B}(K\otimes K)$ is always positive;
if $\rho$ is separable then for every
 positive linear map $\Phi :{\mathcal B}(H)\rightarrow
{\mathcal B}(K)$, the operator $(\Phi\otimes I)\rho$ is always
positive on $K\otimes K$ (or, for every
 positive linear map $\Phi :{\mathcal B}(K)\rightarrow
{\mathcal B}(H)$, the operator $(I\otimes \Phi)\rho$ is always
positive on $H\otimes H$). For the finite dimensional cases, the
converse of the last statement is also true since, due to the
Choi-Jamio{\l}kowski isomorphism, any EW on $H\otimes K$ corresponds
to a positive linear map $\Phi:{\mathcal B}(H)\rightarrow{\mathcal
B}(K)$.  In \cite{Hor}, the following positive map criterion was
established.

{\bf Horodeckis' Theorem.} (\cite[Theorem 2]{Hor}) {\it Let $H$, $K$
be finite dimensional complex Hilbert spaces and $\rho$ be a state
acting on $H\otimes K$. Then $\rho$ is separable if and only if for
any positive linear map $\Phi :{\mathcal B}(H)\rightarrow {\mathcal
B}(K)$, the operator $(\Phi\otimes I)\rho $ is positive on $K\otimes
K$.}

Recently, Hou in \cite{H} improved the above result and established
the elementary operator criterion for infinite dimensional bipartite
systems.

{\bf The elementary operator criterion.} (\cite[Theorem 4.5]{H})
{\it Let $H$, $K$ be complex Hilbert spaces and $\rho$ be a state
acting on $H\otimes K$. Then the following statements are
equivalent.}

(1) {\it $\rho$ is separable;}

(2) {\it $(\Phi\otimes I)\rho \geq 0$ holds for every positive
 elementary operator $\Phi :{\mathcal B}(H)\rightarrow
{\mathcal B}(K)$.}

(3) {\it $(\Phi\otimes I)\rho \geq 0$ holds for every positive
 finite rank elementary operator $\Phi :{\mathcal B}(H)\rightarrow
{\mathcal B}(K)$.}

Therefore, a state $\rho$ on $H\otimes K$ is entangled if and only
if there exists a  positive finite rank elementary operator
$\Phi:{\mathcal B}(H)\rightarrow {\mathcal B}(K)$ which is not
completely positive (briefly, NCP) such that $(\Phi\otimes I)\rho$
is not positive. Thus it is very important and interesting to find
as many as possible positive finite rank  elementary operators  that
are NCP, and then, to apply them to detect the entanglement of
states.

The purpose of this paper is to construct some new indecomposable
positive finite rank elementary operators and apply them to get some
new examples of entangled states that can not be detected by the PPT
criterion and the realignment criterion. Recall that, a positive map
$\Delta$ is said to be decomposable if it is the sum of a completely
positive map $\Delta_1$ and the composition of a completely positive
map $\Delta_2$ and the transpose $\bf T$, i.e.,
$\Delta=\Delta_1+{\bf T}\circ\Delta_2$.

The paper is organized as follows. Section 2 is devoted to giving
some preliminary results on characterizing positive elementary
operators and introducing a concept of the order of finite rank
elementary operators (Definition 2.8). In Section 3,  we give a
simple  necessary and sufficient condition for a pure state to be
separable in bipartite  systems of any dimension in terms of a
special positive elementary operator of order $(2,2)$ (Theorem 3.1).
Then we use a class of positive finite rank elementary operators of
order $(3,3)$
 (see Theorem 3.2) to detect   entanglement of some states (Example
 3.3
and 3.4).  The purpose of Section 4 is to obtain a new class of
positive finite rank elementary operators of order $(4,4)$ that are
not decomposable (Theorem 4.1), and then, apply them to detect the
entanglement of two kinds of states (Example 4.2 and 4.3). Some new
examples of PPT entangled states that can not be detected by the
realignment criterion are obtained. Section 5 is devoted to
discussing the general case. For any $n\geq 3$, a new class of
indecomposable positive finite rank elementary operators of order
$(n,n)$ are constructed (Theorem 5.1) and the entanglement of two
kinds of states are detected (Examples 5.4-5.5). In Section 6, a
short conclusion is given.

Throughout this paper, $H$ and $K$ are complex Hilbert spaces of any
dimension, and $\langle \cdot|\cdot\rangle$ stands for the inner
product in both of them. ${\mathcal B}(H,K)$ (${\mathcal B}(H)$ when
$K=H$) is the Banach space of all (bounded linear) operators from
$H$ into $K$. $A\in{\mathcal B}(H)$ is self-adjoint if $A=A^\dagger$
($A^\dagger$ stands for the adjoint operator of $A$); and $A$ is
positive, denoted by $A\geq 0$, if
 $\langle \psi | A|\psi\rangle\geq 0$ for all $|\psi\rangle\in H$.
For any positive integer $n$, $H^{(n)}$ denotes the direct sum of
$n$ copies of $H$. It is clear that every operator ${\bf
A}\in{\mathcal B}(H^{(n)},K^{(m)})$ can be written in an $m\times n$
operator matrix ${\bf A}=(A_{ij})_{i,j}$ with $A_{ij}\in{\mathcal
B}(H,K)$, $i=1,2,\ldots,m$; $j=1,2,\ldots,n$. Equivalently,
${\mathcal B}(H^{(n)},K^{(m)})$ is often written as ${\mathcal
B}(H,K)\otimes {\mathcal M}_{m\times n}({\mathbb C})$. If $\Phi$ is
a linear map from ${\mathcal B}(H)$ into ${\mathcal B}(K)$, we can
define a linear map $\Phi_n:{\mathcal
B}(H^{(n)})\rightarrow{\mathcal B}(K^{(n)})$ by
$\Phi_n((A_{ij}))=(\Phi(A_{ij}))$. Recall that $\Phi$ is  positive
if $A\in{\mathcal B}(H)$ is positive  implies that $\Phi(A)$ is
positive; $\Phi$ is $n$-positive if $\Phi_n$ is positive; $\Phi$ is
completely positive if $\Phi_n$ is positive for every integer $n>0$.
A linear map $\Phi: {\mathcal B}(H)\rightarrow{\mathcal B}(K)$ is
called an elementary operator if there are two finite sequences
$\{A_i\}^n_{i=1}\subset{\mathcal B}(H,K)$ and
$\{B_i\}^n_{i=1}\subset{\mathcal B}(K,H)$ such that
$\Phi(X)=\sum_{i=1}^n A_iXB_i$ for all $X\in{\mathcal B}(H)$. It was
shown in  \cite{H5} that an elementary operator $\Phi$ is of finite
rank if and only if there exist finite rank operators $A_i, B_i$,
$i=1,2,\cdots ,k$, such that $\Phi(X) =\sum_{i=1}^k A_iXB_i$.

\section{A  characterization of positive elementary operators}

In this section, we give some preliminary results on characterizing
positive elementary operators, which are needed in this paper.

 Before stating the main results in this section, let us recall some
notions from \cite{H4}.
 Let $l$, $k\in\mathbb{N}$ (the set
of all natural numbers),  and let $A_{1},\cdots, A_{k}$, and
$C_{1},\cdots, C_{l}\in {\mathcal B}(H$, $K$). If, for each
$|\psi\rangle\in H^{(m)}$ (the direct sum of $m$ copies of $H$),
there exists an $l\times k$ complex matrix $(\alpha
_{ij}(|\psi\rangle))$ (depending on $|\psi\rangle$) such that
$$
C_{i}^{(m)}|\psi\rangle=\sum _{j=1}^{k}\alpha
_{ij}(|\psi\rangle)A_{j}^{(m)}|\psi\rangle,\qquad i=1,2,\cdots ,l,
$$
we say that $(C_{1},\cdots ,C_{l})$ is an $m$-locally linear
combination of $(A_{1},\cdots ,A_{k})$,
$(\alpha_{ij}(|\psi\rangle))$ is called a {\it local coefficient
matrix} at $|\psi\rangle$.  Furthermore, if  a local coefficient
matrix $(\alpha_{ij}(|\psi\rangle))$
  can be chosen for every $|\psi\rangle\in H^{(m)}$ so that
its operator norm $\|(\alpha _{ij}(|\psi\rangle))\|\leq 1$, we say
that $(C_{1},\cdots ,C_{l})$ is an {\it $m$-contractive locally
linear combination} of $(A_{1},\cdots ,A_{k})$; if there is a matrix
$(\alpha_{ij})$ such that $C_{i}=\sum _{j=1}^{k}\alpha _{ij}A_{j}$
for all $i$, we say that $(C_{1},\cdots ,C_{l})$ is a {\it linear
combination} of $(A_{1},\cdots ,A_{k})$ with coefficient matrix
$(\alpha _{ij})$. We'll omit $``m"$ in the case $m=1$. Sometimes we
also write $\{A_i\}_{i=1}^k$ for $(A_{1},\cdots ,A_{k})$.

The following characterization of $m$-positive elementary operators
was obtained in \cite{H4}, also, see \cite{H}. If $m=1$, we get a
characterization of positive elementary operators.

\textbf{Theorem 2.1.}  {\it Let $\Phi(\cdot) =\sum
_{i=1}^{n}A_{i}(\cdot )B_{i}$ be an elementary operator from
${\mathcal B}(H)$ into ${\mathcal B}(K)$. $\Phi $ is $m$-positive if
and only if there exist $C_{1},\cdots ,C_{k}$ and $D_{1},\cdots
,D_{l}$ in ${\rm span}\{A_{1},\cdots ,A_{n}\}$ with $k+l\leq n$ such
that $(D_{1},\cdots ,D_{l})$ is an $m$-contractive locally linear
combination of $(C_{1},\cdots ,C_{k})$ and
$$
\Phi(X) =\sum _{i=1}^{k}C_{i}(X )C_{i}^{\dagger}-\sum
_{j=1}^{l}D_{j}(X)D_{j}^{\dagger}\eqno(2.1)
$$
for all $X\in{\mathcal B}(H)$. Furthermore, $\Phi$ in Eq.(2.1) is
completely positive if and only if $(D_{1},\cdots ,D_{l})$ is a
linear combination of $(C_{1},\cdots ,C_{k})$ with a contractive
coefficient matrix, and in turn, if and only if there exist $E_1,
E_2, \ldots , E_r$ with $r\leq k$ such that
$$ \Phi=\sum_{i=1}^r E_i(\cdot )E_i^\dagger.$$}

It is obvious that if $\Phi(\cdot) =\sum _{i=1}^{n}A_{i}(\cdot
)B_{i}$ sends  self-adjoint operators  to self-adjoint operators,
then $\Phi$ can be represented in the form  $ \Phi(\cdot) =\sum
_{i=1}^{k}C_{i}(\cdot )C_{i}^{\dagger}-\sum
_{j=1}^{l}D_{j}(\cdot)D_{j}^{\dagger}$ with $C_{1},\cdots ,C_{k}$
and $D_{1},\cdots ,D_{l}$ in span$\{A_{1},\cdots ,A_{n}\}$.
Furthermore, Theorem 2.1 says that,    $\Phi$ is $m$-positive if and
only if $(D_{1},\cdots ,D_{l})$ is an $m$-contractive locally linear
combination of $(C_{1},\cdots ,C_{k})$, and $\Phi$ is completely
positive if and only if $(D_{1},\cdots ,D_{l})$ is a contractive
 linear combination of $(C_{1},\cdots ,C_{k})$.

Since every linear map between matrix algebras is an elementary
operator, by Theorem 2.1, we have

{\bf Corollary 2.2.} {\it Let $H$ and $K$ be finite dimensional
complex Hilbert spaces and let $\Phi :{\mathcal
B}(H)\rightarrow{\mathcal B}(K)$ be a linear map. Then $\Phi$ is
positive if and only if there exist $C_{1},\cdots
,C_{k},D_{1},\cdots ,\\D_{l}\in{\mathcal B}(H,K)$ such that
$\{D_j\}_{j=1}^l$ is a contractive locally linear combination of
$\{C_i\}_{i=1}^k$ and $\Phi
(X)=\sum_{i=1}^kC_iXC_i^\dagger-\sum_{j=1}^lD_jXD_j^\dagger$ for all
$X\in{\mathcal B}(H)$.}

If ${\mathcal L}\subset{\mathcal B}(H,K)$, we will denote by
${\mathcal L}_F$ the subset of all finite-rank operators in
${\mathcal L}$.

By Theorem 2.1, we can get some useful simple conditions to ensure
that a positive elementary operator is  completely positive or not.
The Corollaries 2.3-2.5 below can be found in \cite{H4, H}.

{\bf Corollary 2.3.}  {\it  Assume that $\Phi =\sum
_{i=1}^{k}A_{i}(\cdot )A^\dagger_{i}-\sum _{j=1}^{l}B_{j}(\cdot
)B^\dagger_{j}:{\mathcal B}(H)\rightarrow {\mathcal B}(K)$ is a
positive elementary operator. If any one of the following conditions
holds, then $\Phi $ is completely positive:}

(i) {\it $k\leq 2$.}

(ii) {\it $\dim ({\rm span}\{A_{1},\cdots ,A_{k}\}_{F})\leq 2$.}

(iii) {\it  There exists a vector $|\psi\rangle\in H$  such that
$\{A_{i}|\psi\rangle\}_{i=1}^{k}$  is linearly independent.}

(iv) {\it $\Phi $ is  $[\frac{k+1}{2}]$-positive, where $[t]$ stands
for the integer part of the real number $t$.}

{\bf Corollary 2.4.}  {\it  Assume that $\Phi =\sum
_{i=1}^{k}A_{i}(\cdot )A^\dagger_{i}-\sum _{j=1}^{l}B_{j}(\cdot
)B^\dagger_{j}:{\mathcal B}(H)\rightarrow {\mathcal B}(K)$ is a
positive elementary operator. If  $\Phi $ is not completely
positive, then }

(i) {\it $k\geq 3$,}

(ii) {\it $\dim({\rm span}\{A_{1},\cdots ,A_{k}\}_{F})\geq 3$,}

(iii) {\it For every vector $|\psi\rangle\in H$,
$\{A_{i}|\psi\rangle\}_{i=1}^{k}$  is linearly dependent.}

(iv)  {\it $B_j$ is a finite-rank perturbation of some combination
of $\{A_{i}\}_{i=1}^{k}$ for each  $j=1,2,\ldots ,l$.}

(v) {\it $\Phi _{[\frac{k+1}{2}]}$ is not positive.}

{\bf Corollary 2.5.} {\it Assume that $\Phi =\sum
_{i=1}^{k}A_{i}(\cdot )A^\dagger_{i}-\sum _{j=1}^{l}B_{j}(\cdot
)B^\dagger_{j}:{\mathcal B}(H)\rightarrow {\mathcal B}(K)$ is an
elementary operator. If  there exists some $j$ such that $B_j$ is
not a contractive linear combination of $\{A_{i}\}_{i=1}^{k}$, then
$\Phi $ is not completely positive.}

The following result is easily checked and  useful to us.

 {\bf Proposition 2.6.} {\it Let
$$B_{(t_1,t_2\cdots ,t_n)}=\left(\begin{array}{ccccc} t_1
&-1&-1&\cdots &-1\\
-1 &t_2&-1&\cdots &-1 \\
\vdots &\vdots &\vdots&\ddots &\vdots\\
 -1 &-1&-1&\cdots &t_n
 \end{array}\right)\in M_n({\mathbb C}).$$ If $t_i\geq n-1$ for each $i=1,2, \cdots, n$,
then $B_{(t_1,t_2\cdots ,t_n)}\geq 0$ (that is, $B_{(t_1,t_2\cdots
,t_n)}$ is semi-positive definite); if   $t_i< n-1$ for each $i=1,2,
\cdots, n$, then $B_{(t_1,t_2\cdots ,t_n)}\ngeq 0$. Particularly,
$B_{(t,t,\cdots , t)} \geq 0$ if and only if $t\geq n-1$.}

{\bf Proof.} Assume that $t_i\geq n-1$ for each $i=1,2, \cdots, n$.
Then $t_0=\min\{t_1,t_2,\cdots ,t_n\}\geq n-1$. For any
$|x\rangle=(\xi_1,\xi_2,\ldots ,\xi_n)^T\in {\mathbb C}^n$, we have
 $$\begin{array}{rl} \langle x|B_{(t_1,t_2\cdots ,t_n)}|x\rangle=&t_0\sum_{i=1}^n|\xi_i|^2-2\sum_{i<j}
\xi_i\bar{\xi_j}\\
\geq & t_0\sum_{i=1}^n|\xi_i|^2-2\sum_{i<j}|\xi_i||\xi_j|\\
=&(t_0-n+1)\sum_{i=1}^n|\xi_i|^2+(n-1)\sum_{i=1}^n|\xi_i|^2-2\sum_{i<j}|\xi_i||\xi_j|\\
=&(t_0-n+1)\sum_{i=1}^n|\xi_i|^2+\sum_{i<j}^n(|\xi_i|-|\xi_j|)^2\geq
0.
\end{array}$$
 which
implies that $B_{(t_1,t_2\cdots ,t_n)}\geq 0$. If $t_i< n-1$ for
each $i=1,2, \cdots, n$, then $t_0^\prime=\max\{t_1,t_2,\cdots
,t_n\}<n-1$. Taking $\xi_1=\xi_2=\cdots=\xi_n\not=0$ and let
$|x_0\rangle=(\xi_1,\xi_1,\ldots ,\xi_1)^T$, one gets $\langle
x_0|B_{(t_1,t_2\cdots
,t_n)}|x_0\rangle\leq(t_0^\prime-n+1)n\sum_{i=1}^n|\xi_1|^2<0$. It
follows that $B_t\not\geq 0$, completing the proof. \hfill$\Box$

By using of above results, we can prove the following result.

{\bf Proposition 2.7.} {\it Let $H$ and $K$ be Hilbert spaces and
let $\{|i\rangle\}_{i=1}^n$ and $\{|i^\prime\rangle\}_{i=1}^n$ be
any orthonormal sets of $H$ and $K$, respectively.  Denote
$E_{ji}=|j^\prime\rangle\langle i|\in{\mathcal B}(H,K)$. Let
$\Delta:{\mathcal B}(H)\rightarrow {\mathcal B}(K)$ be defined by
$$\begin{array}{rl} \Delta_{(t_1,t_2,\cdots,t_n)}
(A)=&\sum_{i=1}^{n}t_iE_{ii}AE_{ii}^\dagger-(\sum_{i=1}^{n}E_{ii})A(\sum_{i=1}^{n}E_{ii})^\dagger
\end{array}$$ for all $A\in {\mathcal B}(H)$. If $t_i\geq n$ for each $i=1,2, \cdots, n$,
then $\Delta_{(t_1,t_2,\cdots,t_n)}$ is a completely positive map;
if $t_i< n$ for each $i=1,2, \cdots, n$, then
$\Delta_{(t_1,t_2,\cdots,t_n)}$ is  not a positive map.
Particularly, $\Delta_{(t,t,\cdots,t)}$ is positive if and only if
it is completely positive, and in turn, if and only if $t\geq n$.}

{\bf Proof.} For any unit vector
$|x\rangle=(\xi_1,\xi_2,\cdots,\xi_n,0,0,\cdots)^T\in H$,
 consider the rank-one projection
 $|x\rangle\langle x|$. We have
 $$\begin{array}{rl} \Delta(|x\rangle\langle x|)=&\left(\begin{array}{cccc|ccc}
 (t_1-1)|\xi_1|^2
 &-\xi_1\bar{\xi_2} & \cdots &-\xi_1\bar{\xi_n}&0&0& \cdots \\
-\xi_2\bar{\xi_1} &(t_2-1)|\xi_2|^2 &\cdots
& -\xi_2\bar{\xi_n}&0&0& \cdots \\
\vdots &\vdots &\ddots &\vdots\vdots &\vdots &\ddots &\vdots\\
-\xi_n\bar{\xi_1}
&-\xi_n\bar{\xi_2} & \cdots &(t_n-1)|\xi_n|^2&0&0& \cdots\\
\hline 0
 &0 & \cdots &0&0&0& \cdots \\
0&0&\cdots&0&0&0& \cdots \\
\vdots &\vdots &\ddots &\vdots&\vdots &\vdots &\ddots
\end{array}\right).
\end{array}
\eqno(2.2)$$ If $t_i<n$ for each $i=1,2,\cdots ,n$, taking
$|x\rangle=(1,1,\ldots,1,0,0,\cdots)^T$ in Eq.(2.2) and by
Proposition 2.6, we get $\Delta(|x\rangle\langle x|)\not\geq 0$, and
so $\Delta$ is not positive.

  On the other hand, assume that $t_i\geq n$ for each $i=1,2,\cdots ,n$.
 Since $\sum_{i=1}^{n }E_{ii}=\sum_{i=1}^n \frac{1}{\sqrt{t_i}}({\sqrt{t_i}}E_{ii})$ and
 $\sum_{i=1}^n(\frac{1}{\sqrt{t_i}})^2\leq \sum_{i=1}^n(\frac{1}{\sqrt{n}})^2\leq  1$, $\sum_{i=1}^{n }E_{ii}$ is a
contractive linear combination of $\{{\sqrt{t_1}}E_{11},
{\sqrt{t_2}}E_{22},\cdots ,{\sqrt{t_n}}E_{nn}\}$. By Theorem 2.1,
$\Delta$ is completely positive. \hfill$\square$

For the sake of convenience, we introduce a terminology here.

{\bf Definition 2.8.} Let $\Delta :{\mathcal
B}(H)\rightarrow{\mathcal B}(K)$ be a finite frank elementary
operator. It follows from a characterization of finite rank
elementary operators in \cite{H5} that there exist finite rank
projections $P\in{\mathcal B}(H)$ and $Q\in{\mathcal B}(K)$ such
that $$\Delta (A)=Q\Delta (PAP)Q \mbox{ for all } A\in{\mathcal
B}(H).\eqno(2.3)$$ Let
$$(n,m)=\min\{({\rm rank}(P), {\rm rank}(Q)): (P,Q) \mbox{ satisfies the
equation}\ (2.3)\}.$$ $(n,m)$ is called the order of $\Delta$, and
we say that the elementary operator $\Delta$ is of the order
$(n,m)$.

\section{Positive finite rank elementary operators of order $(2,2)$ and $(3,3)$}

In this section we will construct some positive finite rank
elementary operators of   order $(2,2)$ and $(3,3)$.  Applying  such
positive maps, we give a simple necessary and sufficient condition
for a pure state to be separable. We also use these positive maps to
detect some  entangled mixed states.

Positive elementary operators of order $(2,2)$ are easily
constructed. For example, Let $H$ and $K$ be  Hilbert spaces of
dimension $\geq 2$, and let $\{|i\rangle\}_{i=1}^{\dim H}$ and
$\{|j^\prime\rangle\}_{j=1}^{\dim K}$ be any orthonormal sets of $H$
and $K$, respectively. Let $ \Phi_0:{\mathcal B}(H)\rightarrow
{\mathcal B}(K)$ be defined by
$$\begin{array}{rl} \Phi_0
(A)=&E_{11}AE_{11}^\dagger+E_{22}AE_{22}^\dagger+E_{12}AE_{12}^\dagger\\&+E_{21}AE_{21}^\dagger
-(E_{11}+E_{22})A(E_{11}+E_{22})^\dagger
\end{array}\eqno(3.1)
$$
and
$$\begin{array}{rl} \Psi_0(A)=&(2E_{11}+E_{22})A(2E_{11}+E_{22})^\dagger+E_{12}AE_{12}^\dagger\\
&+E_{21}AE_{21}^\dagger-(E_{11}+E_{22})A(E_{11}+E_{22})^\dagger
\end{array}\eqno(3.2)$$
 for every $A\in{\mathcal B}(H)$, where
$E_{ji}=|j^\prime\rangle\langle i|$. It is obvious that both
$\Phi_0$ and $\Psi_0$ are positive because the map
$$\left(\begin{array}{cc} a_{11}&a_{12}\\ a_{21}&
a_{22}\end{array}\right)\mapsto \left(\begin{array}{cc} a_{22}
&-a_{12}\\ -a_{21}& a_{11}\end{array}\right)$$ and the map
$$\left(\begin{array}{cc} a_{11}&a_{12}\\ a_{21}&
a_{22}\end{array}\right)\mapsto \left(\begin{array}{cc}
3a_{11}+a_{22} &a_{12}\\ a_{21}& a_{11}\end{array}\right)$$ on
$M_2({\mathbb C})$ are positive. A surprising fact is that such
simple positive elementary operator of order $(2,2)$ will be enough
to determine the separability of the pure states.

Let ${\mathcal U}(H)$ (resp. ${\mathcal U}(K)$) be the group of all
unitary operators on $H$ (resp. on $K$). For any map $\Delta
:{\mathcal B}(H)\rightarrow{\mathcal B}(K)$ and any unitary
operators $U\in{\mathcal U}(H)$ and $V\in{\mathcal U}(K)$, the
deduced map  $A\mapsto V^\dagger \Delta(U^\dagger AU)V$ will be
denoted by $\Delta^{U,V}$. The next result give a simple necessary
and sufficient criterion of separability for pure states in
bipartite composite systems of any dimension. This criterion is
easily performed.

{\bf Theorem 3.1.} {\it Let $H$ and $K$ be  Hilbert spaces of
dimension $\geq 2$, and let $\{|i\rangle\}_{i=1}^{\dim H}$ and
$\{|j^\prime\rangle\}_{j=1}^{\dim K}$ be any orthonormal sets of $H$
and $K$, respectively. Let $ \Phi_0 (\Psi_0):{\mathcal
B}(H)\rightarrow {\mathcal B}(K)$ be defined by Eq.(3.1) (Eq.(3.2)).
Then a pure state $\rho$ on $H\otimes K$ is separable if and only if
$$(\Phi_0^{U,V}\otimes I)\rho\geq 0 \ ((\Psi_0^{U,V}\otimes I)\rho\geq 0)$$ holds for all $U\in{\mathcal U}(H)$ and
$V\in{\mathcal U}(K)$. }

{\bf Proof.} If a state $\rho$ is separable, then $(\Phi_0^{U,
V}\otimes I)\rho\geq 0$ ($(\Psi_0^{U, V}\otimes I)\rho\geq 0$) as
$\Phi_0^{U, V}$ ($\Psi_0^{U, V}$) is a positive map.

Conversely, assume that $\rho=|\psi\rangle\langle\psi|$ is an
inseparable pure state. Let
$|\psi\rangle=\sum_{k=1}^{N_{\psi}}\delta_k|k,k^\prime\rangle$ be
the Schmidt decomposition, where $\delta_1\geq\delta_2\geq \cdots
> 0$ with $\sum_{k=1}^{N_\psi}\delta_k^2=1$,  and $\{|k\rangle\}_{k=1}^{N_\psi}$ and
$\{|k^\prime\rangle\}_{k=1}^{N_\psi}$ are orthonormal in $H$ and
$K$, respectively. Thus $\rho=\sum_{k,l
=1}^{N_\psi}\delta_k\delta_{k^\prime}|k,k^\prime\rangle\langle
l,l^\prime|=\sum_{k,l
=1}^{N_\psi}\delta_k\delta_{k^\prime}E_{kl}\otimes E_{k^\prime
l^\prime}$. Since $\rho=|\psi\rangle\langle\psi|$ is inseparable,
the Schmidt number $N_\psi$ of $|\psi\rangle$ is greater than 1 and
hence $\delta_1\geq \delta_2>0$.

Up to unitary equivalence, we may assume that
$\{|k\rangle\}_{k=1}^2=\{|i\rangle\}_{i=1}^2$ and
$\{|k^\prime\rangle\}_{k^\prime=1}^2=\{|j^\prime\rangle\}_{i=1}^2$.
Then, since $\Phi_0(E_{kl})=0$ ($\Psi_0(E_{kl})=0$) whenever $k>2$
or $l>2$, we have
$$\begin{array}{rl}(\Phi_0\otimes
I)\rho=&\sum_{i,j=1}^2\delta_i\delta_j\Phi_0(E_{ij})\otimes
E_{ij}\\
\cong &\left(\begin{array}{cccc}
0&0&0&-\delta_1\delta_2\\
0&\delta_1^2&0&0\\
0&0&\delta_2^2&0\\
-\delta_1\delta_2&0&0&0
\end{array}\right)\oplus 0
\end{array}$$
$$\begin{array}{rl}((\Psi_0\otimes I)\rho=&\sum_{i,j=1}^2\delta_i\delta_j\Psi_0(E_{ij})\otimes
E_{ij}\\ \cong &\left(\begin{array}{cccc}
3\delta_1^2&0&0&\delta_1\delta_2\\
0&\delta_1^2&0&0\\
0&0&\delta_2^2&0\\
\delta_1\delta_2&0&0&0
\end{array}\right)\oplus 0),\end{array}$$
which is clearly not positive. \hfill$\Box$

Now let us consider the positive elementary operators of order
$(3,3)$.

{\bf Theorem 3.2.} {\it Let $H$ and $K$ be  Hilbert spaces of
dimension $\geq 3$, and let $\{|i\rangle\}_{i=1}^3$ and
$\{|j^\prime\rangle\}_{j=1}^3$ be any orthonormal sets of $H$ and
$K$, respectively. Let $\Phi, \Phi^\prime :{\mathcal
B}(H)\rightarrow {\mathcal B}(K)$ be defined by
$$\begin{array}{rl} \Phi
(A)=&2\sum_{i=1}^3E_{ii}AE_{ii}^\dagger+E_{12}AE_{12}^\dagger+E_{23}AE_{23}^\dagger+E_{31}AE_{31}^\dagger\\&-
(\sum_{i=1}^3E_{ii})A(\sum_{i=1}^3E_{ii})^\dagger
\end{array}\eqno(3.3)$$
and
$$\begin{array}{rl} \Phi^\prime
(A)=&2\sum_{i=1}^3E_{ii}AE_{ii}^\dagger+E_{13}AE_{13}^\dagger+E_{21}AE_{21}^\dagger+E_{32}AE_{32}^\dagger\\&-
(\sum_{i=1}^3E_{ii})A(\sum_{i=1}^3E_{ii})^\dagger
\end{array}\eqno(3.3)^\prime$$
 for every $A\in{\mathcal B}(H)$, where $E_{ji}=|j^\prime\rangle\langle i|$.
Then $\Phi$ and $\Phi^\prime$ are indecomposable  positive finite
rank elementary operators of order $(3,3)$.}

{\bf Proof.}  We only give the proof  that $\Phi$ is NCP positive.
$\Phi^\prime$ is dealt with similarly.

It is obvious that $\Phi$ is a finite rank elementary operator of
order $(3,3)$. Also, it is clear from Theorem 2.1 that $\Phi$ is not
completely positive because $\sum_{i=1}^3E_{ii}$ is not a
contractive linear combination of $\{\sqrt{2}E_{11}, \sqrt{2}E_{22},
\sqrt{2}E_{33}, E_{12}, E_{23}, E_{31}\}$. To prove the positivity
of $\Phi$, extend $\{|i\rangle\}_{i=1}^3$ and
$\{|j^\prime\rangle\}_{j=1}^3$ to orthonormal bases
$\{|i\rangle\}_{i=1}^{\dim H}$ and $\{|j^\prime\rangle\}_{j=1}^{\dim
K}$ of $H$ and $K$, respectively. Then every $A\in{\mathcal B}(H)$
has a matrix representation $A=(a_{kl})$ and the map $\Phi$ maps $A$
into
$$\begin{array}{rl} \Phi(A)=&\left(\begin{array}{cccccc}
a_{11}+a_{22}&-a_{12}&-a_{13}&0&0& \cdots \\
-a_{21}&a_{22}+a_{33}&-a_{23}&0&0& \cdots \\
-a_{31}
&-a_{32}& a_{33}+a_{11}&0&0& \cdots\\
0 &0 &0&0&0& \cdots \\
0&0&0&0&0& \cdots \\
\vdots &\vdots &\vdots&\vdots &\vdots &\ddots
\end{array}\right),
\end{array}$$
which is unitarily equivalent to
$$S\oplus0=\left(\begin{array}{ccc}
a_{11}+a_{22}&-a_{12}&-a_{13}\\
-a_{21}&a_{22}+a_{33}&-a_{23} \\
-a_{31} &-a_{32}& a_{33}+a_{11}
\end{array}\right)\oplus 0.$$
By  \cite[Proposition 5.2]{H}, the matrix $S$ is positive. So
$\Phi(A)$ is positive. The  fact that $\Phi$ is not decomposable
will be proved by Example 3.3 or 3.4, completing the proof of the
theorem. \hfill$\Box$

Next we use the positive maps in  Theorem 3.2 to detect some mixed
entangled states. These examples also imply that the positive maps
in Theorem 3.2 are not decomposable since they can recognize some
PPT entangled states.

The states $\rho$ in Example 3.3 was discussed in \cite{JB} and
their entanglement were detected by constructing suitable witnesses.

{\bf Example 3.3.} Let $H$ and $K$ be  Hilbert spaces and let
$\{|i\rangle\}_{i=1}^3$ and $\{|j^\prime\rangle\}_{j=1}^3$ be any
orthonormal sets of $H$ and $K$, respectively. Let
$|\omega\rangle=\frac{1}{\sqrt{3}}(|11'\rangle+|22'\rangle+|33'\rangle)$,
and  define $\rho_1=|\omega\rangle\langle\omega|$,
$\rho_2=\frac{1}{3}(|12'\rangle\langle12'|
+|23'\rangle\langle23'|+|31'\rangle\langle31'|)$ and
$\rho_3=\frac{1}{3}(|13'\rangle\langle13'|
+|21'\rangle\langle21'|+|32'\rangle\langle32'|).$ Let
$\rho=\sum_{i=1}^3q_i\rho_i$ and $\rho_t=(1-t)\rho+t\rho_0$, where
 $q_i\geq 0$ for $i=1,2,3$ with $q_1+q_2+q_3=1$, $t\in[0,1]$,
and $\rho_0$ is a state on $H\otimes K$. By the positive finite rank
elementary operators $\Phi$ and $\Phi^\prime$ defined by Eq.(3.3)
and Eq.(3.3)$^\prime$, respectively, we obtain that,
 for sufficiently small $t$ or for any $\rho_0$ with $(\Phi\otimes I)\rho_0=(\Phi^\prime\otimes I)\rho_0=0$, the following statements are true.

(1) If $q_i<q_1$ for some $i=2,3$, then $\rho_t$ is entangled.

(2) Let  $\rho_0$ be PPT.  Then $\rho_t$ is PPT if and only if
$q_iq_j\geq q_1^2$. Thus, if $0<q_i<q_1<\frac{1}{3}$ and
$\frac{1}{3}<q_j<1$ with $q_iq_j\geq q_1^2$, where $i,j\in\{2,3\}$
and $i\not=j$, then $\rho_t$ is PPT entangled.

In fact, by \cite{HQ}, we need only to check the following:

(1)$^\prime$ if $q_i<q_1$ for some $i=2,3$, then $\rho$ is
entangled;

(2)$^\prime$ $\rho$ is PPT if and only if $q_iq_j\geq q_1^2$. Thus,
if $0<q_i<q_1<\frac{1}{3}$ and $\frac{1}{3}<q_j<1$ with $q_iq_j\geq
q_1^2$, where $i,j\in\{2,3\}$ and $i\not=j$, then $\rho$ is PPT
entangled.

For the map $\Phi$, we have
$$\rho=\frac{1}{3}\left(\begin{array}{ccccccccc}q_1&0&0&0&q_1&0&0&0&q_1\\
0&q_3&0&0&0&0&0&0&0\\
0&0&q_2&0&0&0&0&0&0\\
0&0&0&q_2&0&0&0&0&0\\
q_1&0&0&0&q_1&0&0&0&q_1\\
0&0&0&0&0&q_3&0&0&0\\
0&0&0&0&0&0&q_3&0&0\\
0&0&0&0&0&0&0&q_2&0\\
q_1&0&0&0&q_1&0&0&0&q_1
\end{array}\right)\oplus 0$$and
$$\begin{array}{rl}&3(\Phi\otimes I)(\rho)\\
\cong&{\small \left(\begin{array}{ccccccccc}
q_1+q_3&0&0&0&-q_1&0&0&0&-q_1\\
0&q_2+q_3&0&0&0&0&0&0&0\\
0&0&q_1+q_2&0&0&0&0&0&0\\
0&0&0&q_1+q_2&0&0&0&0&0\\
-q_1&0&0&0&q_1+q_3&0&0&0&-q_1\\
0&0&0&0&0&q_2+q_3&0&0&0\\
0&0&0&0&0&0&q_2+q_3&0&0\\
0&0&0&0&0&0&0&q_1+q_2&0\\
-q_1&0&0&0&-q_1&0&0&0&q_1+q_3\end{array}\right)}\oplus 0\\
\cong&A_1\oplus B_1\oplus C_1\oplus 0,\end{array}$$ where $\cong$
means ``be unitarily equivalent to",
$$A_1=
 \left(\begin{array}{ccc}
q_1+q_3&-q_1&-q_1\\
-q_1&q_1+q_3&-q_1\\
-q_1&-q_1&q_1+q_3\end{array}\right),\  B_1=
\left(\begin{array}{ccc}
q_1+q_2&0&0\\
0&q_1+q_2&0\\
0&0&q_1+q_2\end{array}\right)$$ and $$C_1=  \left(\begin{array}{ccc}
q_2+q_3&0&0\\
0&q_2+q_3&0\\
0&0&q_2+q_3\end{array}\right).$$It is obvious that $B_1,C_1\geq0$.
For $A_1$, by Proposition 2.6, we have $A_1\ngeqslant 0$ if
$q_3<q_1$. It follows from the elementary operator criterion that
$\rho$ is entangled if $q_3<q_1$. Moreover, it is easily checked
that $\rho$ is PPT if and only if $q_2q_3\geq q_1^2$. Thus we obtain
that $\rho$ is PPT entangled if $0<q_3<q_1<\frac{1}{3}$ and
$\frac{1}{3}\leq q_2<1$ with $q_2q_3\geq q_1^2$.

Similarly, by using of the map $\Phi^\prime$, one gets the other
half of the assertions (1)$^\prime$-(2)$^\prime$.

The states $\rho_t$ in the next example were introduced in \cite{HQ}
firstly.

{\bf Example 3.4.} Let $H$ and $K$ be complex Hilbert spaces and let
$\{|i\rangle\}_{i=1}^{\dim H}$ and $\{|j^\prime\rangle\}_{j=1}^{\dim
K}$ be any orthonormal bases of $H$ and $K$, respectively.  Let
$$|\omega_{1}\rangle=\frac{1}{\sqrt{3}}(|11'\rangle+|22'\rangle+|33'\rangle) \ \
{\rm and} \ \
|\omega_{2}\rangle=\frac{1}{\sqrt{3}}(|12'\rangle+|23'\rangle+|31'\rangle).$$
Define $\rho_1=|\omega_1\rangle\langle\omega_1|$,
$\rho_2=|\omega_2\rangle\langle\omega_2|$ and
$\rho_3=\frac{1}{3}(|13'\rangle\langle13'|
+|21'\rangle\langle21'|+|32'\rangle\langle32'|).$ Let
$\rho=\sum_{i=1}^3q_i\rho_i$ and $\rho_t=(1-t)\rho+t\rho_0$, where
$q_i\geq 0$ for $i=1,2,3$ with $q_1+q_2+q_3=1$, $t\in[0,1]$, and
$\rho_0$ is a state on $H\otimes K$.

Hou and Qi in \cite{HQ} proved that, if $q_2<\frac{5}{7}q_1$ or
$q_1<\frac{5}{7}q_2$, then,  for sufficient small $t$, $\rho_t$ is
entangled; if $q_2<\frac{5}{7}q_1$ or $q_1<\frac{5}{7}q_2$, and if
$q_1q_2q_3\geq q_1^3+q_2^3$, then $\rho_t$ is PPT entangled whenever
$\rho_0$ is. Now, by using of the positive finite rank elementary
operators $\Phi$ and $\Phi^\prime$ in Theorem 3.2, we can give a
finer result. In fact, for sufficient small $t$, or for $\rho_0$
with $(\Phi\otimes I)\rho_0=(\Phi^\prime\otimes I)\rho_0=0$ (for
example, taking
$\rho_0=\sum\limits_{i=4}^{\infty}p_i|i\rangle\langle
i^\prime|\otimes|i\rangle\langle i^\prime|,\ p_i\geq0,\
\sum\limits_{i=4}^{\infty}p_i=1$), the following statements are
true.

(1) If $q_1\not=q_2$ or $q_1=q_2>q_3$,  then $\rho_t$ is entangled;

(2) Let $\rho_0$ be PPT. Then $\rho_t$ is PPT if and only if
$q_1q_2q_3\geq q_1^3+q_2^3$. Particularly, if $q_j=2q_i$ and
$\frac{9}{2}q_j\leq q_3$, where $i,j\in\{1,2\}$ and $i\not=j$, then
$\rho_t$ is PPT entangled.

Still, we need only consider $\rho$ and check the following:

(1)$^\prime$ If $q_1\not=q_2$ or $q_1=q_2>q_3$,  then $\rho$ is
entangled;

(2)$^\prime$ $\rho$ is PPT if and only if $q_1q_2q_3\geq
q_1^3+q_2^3$. Particularly, if $q_j=2q_i$ and $\frac{9}{2}q_j\leq
q_3$, where $i,j\in\{1,2\}$ and $i\not=j$, then $\rho$ is PPT
entangled.

For $\rho=q_1\rho_1+q_2\rho_2+q_3\rho_3$, it is obvious that
$$\rho=\frac{1}{3}\left(\begin{array}{ccccccccc}q_1&0&0&0&q_1&0&0&0&q_1\\
0&q_3&0&0&0&0&0&0&0\\
0&0&q_2&q_2&0&0&0&q_2&0\\
0&0&q_2&q_2&0&0&0&q_2&0\\
q_1&0&0&0&q_1&0&0&0&q_1\\
0&0&0&0&0&q_3&0&0&0\\
0&0&0&0&0&0&q_3&0&0\\
0&0&q_2&q_2&0&0&0&q_2&0\\
q_1&0&0&0&q_1&0&0&0&q_1
\end{array}\right)\oplus 0. $$ Note that
$$\begin{array}{rl}&3(\Phi\otimes I)(\rho)\\
\cong&{\small \left(\begin{array}{ccccccccc}
q_1+q_3&0&0&0&-q_1&0&0&0&-q_1\\
0&q_2+q_3&0&0&0&0&0&0&0\\
0&0&q_1+q_2&-q_2&0&0&0&-q_2&0\\
0&0&-q_2&q_1+q_2&0&0&0&-q_2&0\\
-q_1&0&0&0&q_1+q_3&0&0&0&-q_1\\
0&0&0&0&0&q_2+q_3&0&0&0\\
0&0&0&0&0&0&q_2+q_3&0&0\\
0&0&-q_2&-q_2&0&0&0&q_1+q_2&0\\
-q_1&0&0&0&-q_1&0&0&0&q_1+q_3\end{array}\right)}\oplus
0,\end{array}$$which is unitarily equivalent to the operator
$A\oplus B\oplus C\oplus 0$, where
$$A= \left(\begin{array}{ccc}
q_1+q_3&-q_1&-q_1\\
-q_1&q_1+q_3&-q_1\\
-q_1&-q_1&q_1+q_3\end{array}\right),\  B=  \left(\begin{array}{ccc}
q_1+q_2&-q_2&-q_2\\
-q_2&q_1+q_2&-q_2\\
-q_2&-q_2&q_1+q_2\end{array}\right)$$ and $$C=
 \left(\begin{array}{ccc}
q_2+q_3&0&0\\
0&q_2+q_3&0\\
0&0&q_2+q_3\end{array}\right)\geq 0.$$ For the matrices $A$ and $B$,
by Proposition 2.6, we get that $A\ngeq 0$ if $q_3<q_1$ and $B\ngeq
0$ if $q_1<q_2$. So $(\Phi\otimes I)(\rho)$ is not positive if
$q_3<q_1$ or $q_1<q_2$. It follows from the elementary operator
criterion that $\rho$ is entangled if $q_3<q_1$ or $q_1<q_2$. Note
that $\rho$ is PPT if and only if $q_1q_2q_3\geq q_1^3+q_2^3$. Thus
particularly we obtain that $\rho$ is PPT entangled if $q_2=2q_1$
and $\frac{9}{2}q_1\leq q_3$.

Similarly, by applying the map $\Phi^\prime$, one can get that the
other half of the assertions (1)$^\prime$-(2)$^\prime$ is true.

\section{Positive finite rank elementary operators of order $(4,4)$}

In this section we construct a class of positive finite rank
elementary operators of order $(4,4)$. The following is our main
result.

{\bf Theorem 4.1.} {\it Let $H$ and $K$ be  Hilbert spaces of
dimension greater than 3 and let $\{|i\rangle\}_{i=1}^4$ and
$\{|j^\prime\rangle\}_{j=1}^4$ be any orthonormal sets of $H$ and
$K$, respectively. Let $\Phi, \Phi^\prime,
\Phi^{\prime\prime}:{\mathcal B}(H)\rightarrow {\mathcal B}(K)$ be
defined by
$$\begin{array}{rl} \Phi
(A)=&3\sum_{i=1}^4E_{ii}AE_{ii}^\dagger+E_{12}AE_{12}^\dagger+E_{23}AE_{23}^\dagger
+E_{34}AE_{34}^\dagger+E_{41}AE_{41}^\dagger\\&-
(\sum_{i=1}^4E_{ii})A(\sum_{i=1}^4E_{ii})^\dagger,
\end{array}\eqno(4.1)$$
$$\begin{array}{rl} \Phi^\prime
(A)=&3\sum_{i=1}^4E_{ii}AE_{ii}^\dagger+E_{13}AE_{13}^\dagger+E_{24}AE_{24}^\dagger
+E_{31}AE_{31}^\dagger+E_{42}AE_{42}^\dagger\\&-
(\sum_{i=1}^4E_{ii})A(\sum_{i=1}^4E_{ii})^\dagger
\end{array}\eqno(4.1)^\prime$$and
$$\begin{array}{rl} \Phi^{\prime\prime}
(A)=&3\sum_{i=1}^4E_{ii}AE_{ii}^\dagger+E_{14}AE_{14}^\dagger+E_{21}AE_{21}^\dagger
+E_{32}AE_{32}^\dagger+E_{43}AE_{43}^\dagger\\&-
(\sum_{i=1}^4E_{ii})A(\sum_{i=1}^4E_{ii})^\dagger
\end{array}\eqno(4.1)^{\prime\prime}$$ for every $A\in{\mathcal B}(H)$, where $E_{ji}=|j^\prime\rangle\langle i|$.
Then $\Phi, \Phi^\prime, \Phi^{\prime\prime}$ are positive finite
rank elementary operators that are not completely positive.
Moreover, $\Phi$ and $\Phi^{\prime\prime}$ are indecomposable.}

{\bf Proof.} Still, we only prove that $\Phi$ is positivity but not
completely positive.

It is clear from Theorem 2.1 that $\Phi$ is not completely positive
because $\sum_{i=1}^4E_{ii}$ is not a contractive linear combination
of $\{\sqrt{3}E_{11},\ldots ,\sqrt{3}E_{44}$, $E_{12}$, $E_{23}$,
$E_{34}$,  $E_{41}\}$. We will show that $\Phi$ is positive. Extend
$\{|i\rangle\}_{i=1}^4$ and $\{|j^\prime\rangle\}_{j=1}^4$ to
orthonormal bases $\{|i\rangle\}_{i=1}^{\dim H}$ and
$\{|j^\prime\rangle\}_{j=1}^{\dim K}$ of $H$ and $K$, respectively.
Then every $A\in{\mathcal B}(H)$ has a matrix representation
$A=(a_{kl})$. Obviously, $\Phi$ maps $A=(a_{kl})$ to the matrix
$$\Phi(A)=\left( \begin{array}{ccccccc} 2a_{11}+a_{22}
&-a_{12}&-a_{13}&-a_{14}&0&\cdots\\
-a_{21} &2a_{22}+a_{33} &-a_{23}&-a_{24}&0&\cdots \\
-a_{31} &-a_{32} & 2a_{33}+a_{44}&-a_{34}&0&\cdots \\
-a_{41} &-a_{42} &-a_{43}&
2a_{44}+a_{11}&0&\cdots\\
0&0&0&0&0&\cdots\\
\vdots&\vdots&\vdots&\vdots&\vdots&\ddots\end{array}\right).$$ Take
any unit vector $|x\rangle=(x_1,x_2,x_3 ,x_4, x_5,\cdots)^T\in H$
and
 consider the rank-one projection
 $|x\rangle\langle x|$. Obviously, $\Phi$ is positive if and only if $\Phi(|x\rangle\langle x|)\geq 0$
 holds for all unit vector $x\in H$. Since
 $$\begin{array}{rl} \Phi(|x\rangle\langle x|)=\left(\begin{array}{ccccccc}
 2|x_1|^2+|x_2|^2 & -x_1\bar{x_2} & -x_1\bar{x_3} & -x_1\bar{x_4}&0&\cdots\\
-x_2\bar{x_1} & 2|x_2|^2+|x_3|^2 & -x_2\bar{x_3} & -x_2\bar{x_4}&0&\cdots\\
-x_3\bar{x_1} & -x_3\bar{x_2} & 2|x_3|^2+|x_4|^2  & -x_3\bar{x_4}&0&\cdots\\
-x_4\bar{x_1} &-x_4\bar{x_2} &-x_4\bar{x_3}
&2|x_4|^2+|x_1|^2&0&\cdots\\
0&0&0&0&0&\cdots\\
\vdots&\vdots&\vdots&\vdots&\vdots&\ddots\end{array}\right).
\end{array}$$
we see that $\Phi(|x\rangle\langle x|)\geq 0$ if and only if
$$M(x)=\left(\begin{array}{cccc}
 2|x_1|^2+|x_2|^2 & -x_1\bar{x_2} & -x_1\bar{x_3} & -x_1\bar{x_4}\\
-x_2\bar{x_1} & 2|x_2|^2+|x_3|^2 & -x_2\bar{x_3} & -x_2\bar{x_4}\\
-x_3\bar{x_1} & -x_3\bar{x_2} & 2|x_3|^2+|x_4|^2  & -x_3\bar{x_4}\\
-x_4\bar{x_1} &-x_4\bar{x_2} &-x_4\bar{x_3}
&2|x_4|^2+|x_1|^2\end{array}\right)\geq 0.$$ It follows from
Proposition 2.6  that all the principal minor determinants with
order less than 4 of matrix $M(x)$ are semi-positive definite. So,
to prove the positivity of $M(x)$, we need only to show that $\det(
M(x))\geq 0$. Writing $x_i=r_ie^{i\theta_i}$, $i=1,2,3,4$, we have
$$M(x)=U\left(\begin{array}{cccc}
 2r_1^2+r_2^2 & -r_1r_2 & -r_1r_3 & -r_1r_4\\
-r_1r_2 & 2r_2^2+r_3^2 & -r_2r_3 & -r_2r_4\\
-r_1r_3 & -r_2r_3 & 2r_3^2+r_4^2  & -r_3r_4\\
-r_1r_4 &-r_2r_4 &-r_3r_4
&2r_4^2+r_1^2\end{array}\right)U^\dagger,$$ where
$$U=\left(\begin{array}{cccc}
 e^{i\theta_1} &0& 0 & 0\\
0& e^{i\theta_2} &0& 0\\
0 & 0 & e^{i\theta_3}  & 0\\
0 &0&0&e^{i\theta_4} \end{array}\right)$$is a unitary matrix. It
follows that $\Phi$ is positive if and only if the determinant
$$f(r_1,r_2,r_3,r_4)=\begin{vmatrix}
 2r_1^2+r_2^2 & -r_1r_2 & -r_1r_3 & -r_1r_4\\
-r_1r_2 & 2r_2^2+r_3^2 & -r_2r_3 & -r_2r_4\\
-r_1r_3 & -r_2r_3 & 2r_3^2+r_4^2  & -r_3r_4\\
-r_1r_4 &-r_2r_4 &-r_3r_4 &2r_4^2+r_1^2\end{vmatrix}\geq0$$ holds
for all $0\leq r_1,r_2,r_3,r_4\leq 1$ with
$r_1^2+r_2^2+r_3^2+r_4^2=1$. This is the case since, by a
computation, $\min f(r_1,r_2,r_3,r_4)=0$ (also, refer to the proof
of Theorem 5.1). So $\Phi$ is
 positive, as desired.

Similarly, one can show that $\Phi^\prime$ and $\Phi^{\prime\prime}$
are positive but not completely positive. The fact that $\Phi$ and
$\Phi^{\prime\prime}$ are indecomposable will be
 illustrated by Example 4.2 or 4.3 below. \hfill$\Box$

Now let us give some examples.

The entanglement of the states $\rho$ in Example 4.2 were studied in
\cite{JB} by constructing suitable witnesses. We detect them by the
positive maps obtained in Theorem 4.1. In addition, we also discuss
the question when these states are entangled but cannot be
recognized by the PPT criterion and the realignment criterion.

{\bf Example 4.2.} Let $H$ and $K$ be  Hilbert spaces of dimension
$\geq 4$, and let $\{|i\rangle\}_{i=1}^4$ and
$\{|j^\prime\rangle\}_{j=1}^4$ be any orthonormal sets of $H$ and
$K$, respectively. Let
$|\omega\rangle=\frac{1}{2}(|11'\rangle+|22'\rangle+|33'\rangle+|44'\rangle).$
Define $\rho_1=|\omega\rangle\langle\omega|$,
$\rho_2=\frac{1}{4}(|12'\rangle\langle12'|+|23'\rangle\langle23'|
+|34'\rangle\langle34'|+|41'\rangle\langle41'|),$
$\rho_3=\frac{1}{4}(|13'\rangle\langle13'|
+|24'\rangle\langle24'|+|31'\rangle\langle31'|+|42'\rangle\langle42'|)$
and $\rho_4=\frac{1}{4}(|14'\rangle\langle14'|
+|21'\rangle\langle21'|+|32'\rangle\langle32'|+|43'\rangle\langle43'|).$
Let $\rho=\sum_{i=1}^4q_i\rho_i$ and $\rho_t=(1-t)\rho+t\rho_0$,
where
 $q_i\geq 0$ for $i=1,2,3,4$ with $q_1+q_2+q_3+q_4=1$, $t\in[0,1]$,
and $\rho_0$ is a state on $H\otimes K$. Then for sufficiently small
$t$, or for $\rho_0$ with $(\Phi\otimes I)\rho_0=(\Phi^\prime\otimes
I)\rho_0=(\Phi^{\prime\prime}\otimes I)\rho_0=0$, the following
statements are true.

(1) If $q_i<q_1$ for some $i=2,3,4$, then $\rho_t$ is entangled.

(2)   Let $\rho_0$ be PPT. Then $\rho_t$ is PPT if and only if
$q_2q_4\geq q_1^2$ and $q_3\geq q_1^2$. Thus,  if
$0<q_i<q_1<\frac{1}{4}$, $\frac{1}{4}\leq q_j<1$ with $q_iq_j\geq
q_1^2$ and $0<q_1\leq q_3<1$, where $i,j\in\{2,4\}$ and $i\not=j$,
then $\rho_t$ is PPT entangled;

(3) if $\rho_0$ is PPT, and if $q_1\leq\frac{1}{7}$,
$q_i=\frac{1}{2}q_1$, $q_j=\frac{1}{2}$ and $q_3=\frac{1}{2}-3q_i$,
where $i,j\in\{2,4\}$ and $i\not=j$,  then $\rho_t$ is PPT entangled
but can not be detected by the realignment criterion.

We need only check $\rho$.

 Denote by $F_{k,l}$ the unit matrix with
$(k,l)$-entry 1 and others 0. For $\rho=\sum_{i=1}^4q_i\rho_i$, we
have
$$\begin{array}{rl}\rho=&\frac{1}{4}{\rm
diag}(q_1,q_4,q_3,q_2,q_2,q_1,q_4,q_3,q_3,q_2,q_1,q_4,q_4,q_3,q_2,q_1)\\
&+\frac{q_1}{4}(F_{1,6}+F_{1,11}+F_{1,16}+F_{6,1}+F_{6,11}+F_{6,16}\\
&+F_{11,1}+F_{11,6}+F_{11,16}+F_{16,1}+F_{16,6}+F_{16,11})
\end{array}$$and
$$\begin{array}{rl}&4(\Phi\otimes I)(\rho)\\
=&{\rm
diag}(2q_1+q_4,2q_4+q_3,2q_3+q_2,2q_2+q_1,2q_2+q_1,2q_1+q_4,2q_4+q_3,\\
&2q_3+q_2,2q_3+q_2,2q_2+q_1,2q_1+q_4,2q_4+q_3,2q_4+q_3,2q_3+q_2,2q_2+q_1,2q_1+q_4)\\
&-q_1(F_{1,6}+F_{1,11}+F_{1,16}+F_{6,1}+F_{6,11}+F_{6,16}\\
&+F_{11,1}+F_{11,6}+F_{11,16}+F_{16,1}+F_{16,6}+F_{16,11}),\end{array}$$
which is unitarily equivalent to
$$\begin{array}{rl}&\left(\begin{array}{cccc}
2q_1+q_4&-q_1&-q_1&-q_1\\
-q_1&2q_1+q_4&-q_1&-q_1\\
-q_1&-q_1&2q_1+q_4&-q_1\\
-q_1&-q_1&-q_1&2q_1+q_4\end{array}\right) \oplus ( 2q_4+q_3)I_4 \\
&\oplus (2q_3+q_2)I_4\oplus (2q_2+q_1)I_4\oplus 0.\end{array}
$$
Hence, by Proposition 2.6, we get that $(\Phi\otimes
I)(\rho)\ngeqslant 0$ if  $q_4<q_1$, which implies that $\rho$ is
entangled if $q_4<q_1$.

Note that $$\rho\   \mbox{ \rm is PPT if and only if} \ q_2q_4\geq
q_1^2\ \mbox{\rm and} \ q_3\geq q_1.\eqno(4.2)$$ Thus we obtain that
$\rho$ is PPT entangled if $0<q_4<q_1<\frac{1}{4}$, $\frac{1}{4}\leq
q_2<1$ with $q_2q_4\geq q_1^2$ and $0<q_1\leq q_3<1$. This reveals
that the positive map $\Phi$ can recognize some PPT entangled states
and hence is not decomposable.

The realignment matrix of $\rho$ is
$$\begin{array}{rl}\rho^R\cong&\frac{1}{4}{\rm
diag}(q_1,q_1,q_1,q_1,q_1,q_1,q_1,q_1,q_1,q_1,q_1,q_1,q_1,q_1,q_1,q_1)\\
&+\frac{q_4}{4}(F_{1,6}+F_{6,11}+F_{11,16}+F_{16,1})+\frac{q_3}{4}(F_{1,11}+F_{6,16}+F_{11,1}+F_{16,6})
\\&+\frac{q_2}{4}(F_{1,16}+F_{6,1}+F_{11,6}+F_{16,11})\\
\cong &\frac{1}{4}\left(\begin{array}{cccc}
q_1&q_4&q_3&q_2\\
q_2&q_1&q_4&q_3\\
q_3&q_2&q_1&q_4\\
q_4&q_3&q_2&q_1\end{array}\right)\oplus \frac{1}{4}q_1I_{12}\oplus
0=A\oplus\frac{1}{4}q_1I_{12}  \oplus 0.
\end{array}$$
Thus $\|\rho^R\|_1=\|A\|_1+3q_1$. By computation, we have that
$$\begin{array}{rl}\|A\|_1=&\frac{3}{4}\sqrt{\sum_{i=1}^4q_i^2-q_1q_2-q_2q_3-q_3q_4-q_1q_4}\\
&+\frac{1}{4}\sqrt{\sum_{i=1}^4q_i^2+3(q_1q_2+q_2q_3+q_3q_4+q_1q_4)}.\end{array}\eqno(4.3)$$
it follows from Eqs.(4.2)-(4.3) that  the PPT criterion and the
realignment criterion are independent each other. It is also easy to
construct entangled states that can not be recognized by the PPT
criterion and the realignment criterion. In fact, we have that
$\|\rho^R\|_1<1$ if $q_1\leq\frac{1}{7}$, $q_4=\frac{1}{2}q_1$,
$q_2=\frac{1}{2}$ and $q_3=\frac{1}{2}-3q_4$. For example,
$\|\rho^R\|_1\doteq 0.9411<1$ if $q_1=\frac{1}{7}$,
$q_4=\frac{1}{14}$, $q_2=\frac{1}{2}$ and $q_3=\frac{2}{7}$. Hence,
in this case,  the state $\rho$ is PPT and cannot be detected by the
realignment criterion. However it is entangled and can be recognized
by the positive map $\Phi$ in Theorem 4.1.

Similarly, by applying the map $\Phi^{\prime\prime}$, we have that
$\rho$ is entangled if $q_2<q_1$, and, $\rho$ is PPT entangled if
$0<q_2<q_1<\frac{1}{4}$, $\frac{1}{4}\leq q_4<1$ with $q_2q_4\geq
q_1^2$ and $0<q_1\leq q_3<1$. Thus, $\Phi^{\prime\prime}$ is
indecomposable, too. Furthermore, if $q_1\leq\frac{1}{7}$,
$q_2=\frac{1}{2}q_1$, $q_4=\frac{1}{2}$ and $q_3=\frac{1}{2}-3q_2$,
then $\rho$ is PPT entangled that cannot be detected by the
realignment criterion. However, it can be detected by the positive
map $\Phi^{\prime\prime}$ in Theorem 4.1.

By applying the map $\Phi^{\prime}$, we see that $\rho$ is entangled
if $q_3<q_1$. However, one should be careful that,  in this case,
$\rho$ is not PPT. This means that we can not use $\rho$ to check
whether or not $\Phi^\prime$ is decomposable.

The  following  example is new.

{\bf Example 4.3.} Let $H$ and $K$ be complex Hilbert spaces of
dimension $\geq4$ and let $\{|i\rangle\}_{i=1}^{\dim H}$ and
$\{|j^\prime\rangle\}_{j=1}^{\dim K}$ be any orthonormal bases of
$H$ and $K$, respectively.  Let
$$|\omega_{1}\rangle=\frac{1}{2}(|11'\rangle+|22'\rangle+|33'\rangle+|44'\rangle) \ \
{\rm and} \ \
|\omega_{2}\rangle=\frac{1}{2}(|12'\rangle+|23'\rangle+|34'\rangle+|41'\rangle).$$
Define $\rho_1=|\omega_1\rangle\langle\omega_1|$,
$\rho_2=|\omega_2\rangle\langle\omega_2|$,
$\rho_3=\frac{1}{4}(|13'\rangle\langle13'|
+|24\rangle\langle24'|+|31'\rangle\langle31'|+|42'\rangle\langle42'|)$
and $\rho_4=\frac{1}{4}(|14'\rangle\langle14'|
+|21'\rangle\langle21'|+|32'\rangle\langle32'|+|43'\rangle\langle43'|)$.
Let $\rho=\sum_{i=1}^4q_i\rho_i$ and $\rho_t=(1-t)\rho+t\rho_0$,
where $q_i\geq 0$ for $i=1,2,3,4$ with $q_1+q_2+q_3+q_4=1$,
$t\in[0,1]$, and $\rho_0$ is a state on $H\otimes K$. By using of
the positive finite rank elementary operators $\Phi$, $\Phi^\prime$
and $\Phi^{\prime\prime}$ in Theorem 4.1, we get that, for
sufficient small $t$ or for any $\rho_0$ with $(\Phi\otimes
I)\rho_0=(\Phi^\prime\otimes I)\rho_0=(\Phi^{\prime\prime}\otimes
I)\rho_0=0$, the followings are true.

(1) If $q_1\not=q_2$ or $q_1=q_2>q_i$ for some $i\in\{3,4\}$, then
$\rho_t$ is entangled.

(2) Let $\rho_0$ be PPT. Then $\rho_t$ is PPT if and only if
$q_1(q_1q_3^2-q_2^2q_3-q_1^3)\geq q_2^2(q_1q_3-q_2^2)\geq0$ and
$q_2(q_2q_4^2-q_1^2q_4-q_2^3)\geq q_1^2(q_2q_4-q_1^2)\geq0$. Hence,
if, in addition, $q_1\not=q_2$ or $q_1=q_2>q_i$ for some
$i\in\{3,4\}$, then $\rho_t$ is PPT entangled.

(3) If $\rho_0$ is separable, and if
$\frac{1}{2}q_i=q_j\leq\frac{1}{15}$ and $q_3=q_4$, where
$i,j\in\{1,2\}$ and $i\not=j$, then $\rho_t$ is PPT entangled that
cannot be detected by the realignment criterion.

We need only deal with $\rho$.

For $\rho=q_1\rho_1+q_2\rho_2+q_3\rho_3+4\rho_4$, it is obvious that
$$\begin{array}{rl}\rho=
&\frac{q_1}{4}(F_{1,1}+F_{1,6}+F_{1,11}+F_{1,16}+F_{6,1}+F_{6,6}+F_{6,11}+F_{6,16}\\
&+F_{11,1}+F_{11,6}+F_{11,11}+F_{11,16}+F_{16,1}+F_{16,6}+F_{16,11}+F_{16,16})\\
&+\frac{q_2}{4}(F_{4,4}+F_{4,5}+F_{4,10}+F_{4,15}+F_{5,4}+F_{5,5}+F_{5,10}+F_{5,15}\\
&+F_{10,4}+F_{10,5}+F_{10,10}+F_{10,15}+F_{15,4}+F_{15,5}+F_{15,10}+F_{15,15})\\
&+\frac{q_3}{4}(F_{3,3}+F_{8,8}+F_{9,9}+F_{14,14})
+\frac{q_4}{4}(F_{2,2}+F_{7,7}+F_{12,12}+F_{13,13}).\end{array}
$$
Note that
$${\small\begin{array}{rl} 4(\Phi\otimes I)(\rho)=
&{\rm diag}(2q_1+q_4,q_3+2q_4,q_2+2q_3,q_1+2q_2,q_1+2q_2,2q_1+q_4,q_3+2q_4,q_2+2q_3,\\
&q_2+2q_3,q_1+2q_2,2q_1+q_4,q_3+2q_4,q_3+2q_4,q_2+2q_3,q_1+2q_2,2q_1+q_4)\\
&-q_1(F_{1,6}+F_{1,11}+F_{1,16}+F_{6,1}+F_{6,11}+F_{6,16}\\
&+F_{11,1}+F_{11,6}+F_{11,16}+F_{16,1}+F_{16,6}+F_{16,11})\\
&-q_2(F_{4,5}+F_{4,10}+F_{4,15}+F_{5,4}+F_{5,10}+F_{5,15}\\
&+F_{10,4}+F_{10,5}+F_{10,15}+F_{15,4}+F_{15,5}+F_{15,10}),\end{array}}
$$
which is unitarily equivalent to the operator $A\oplus B\oplus
C\oplus D\oplus 0$, where
$${\small A=\left(\begin{array}{cccc}
2q_1+q_4&-q_1&-q_1&-q_1\\
-q_1&2q_1+q_4&-q_1&-q_1\\
-q_1&-q_1&2q_1+q_4&-q_1\\
-q_1&-q_1&-q_1&2q_1+q_4\end{array}\right), B=
 \left(\begin{array}{cccc}
q_1+2q_2&-q_2&-q_2&-q_2\\
-q_2&q_1+2q_2&-q_2&-q_2\\
-q_2&-q_2&q_1+2q_2&-q_2\\
-q_2&-q_2&-q_2&q_1+2q_2\end{array}\right)}$$ and $${\small C=
\left(\begin{array}{cccc}
q_2+2q_3&0&0&0\\
0&q_2+2q_3&0&0\\
0&0&q_2+2q_3&0\\
0&0&0&q_2+2q_3\end{array}\right), D=  \left(\begin{array}{cccc}
q_3+2q_4&0&0&0\\
0&q_3+2q_4&0&0\\
0&0&q_3+2q_4&0\\
0&0&0&q_3+2q_4\end{array}\right)}.$$ It is clear that $C,D\geq0$.
For the matrices $A$ and $B$, by Proposition 2.6, we get that
$A\geq0$ if and only if $q_4\geq q_1$ and $B\geq 0$ if and only if
$q_1\geq q_2$. So $(\Phi\otimes I)(\rho)$ is not positive if
$q_4<q_1$ or $q_1<q_2$. It follows from the elementary operator
criterion that $\rho$ is entangled if $q_4<q_1$ or $q_1<q_2$.

Next, consider the positive partial transpose of $\rho$. It is clear
that
$$\begin{array}{rl}\rho^{T_1}\cong
&\frac{q_1}{4}(F_{1,1}+F_{2,5}+F_{3,9}+F_{4,13}+F_{5,2}+F_{6,6}+F_{7,10}+F_{8,14}\\
&+F_{9,3}+F_{10,7}+F_{11,11}+F_{12,15}+F_{13,4}+F_{14,8}+F_{15,12}+F_{16,16})\\
&+\frac{q_2}{4}(F_{1,8}+F_{2,12}+F_{3,16}+F_{4,4}+F_{5,5}+F_{6,9}+F_{7,13}+F_{8,1}\\
&+F_{9,6}+F_{10,10}+F_{11,14}+F_{12,2}+F_{13,7}+F_{14,11}+F_{15,15}+F_{16,3})\\
&+\frac{q_3}{4}(F_{3,3}+F_{8,8}+F_{9,9}+F_{14,14})
+\frac{q_4}{4}(F_{2,2}+F_{7,7}+F_{12,12}+F_{13,13})\\
\cong &A_1\oplus B_1\oplus C_1\oplus D_1\oplus0,\end{array}
$$
where
$$A_1={\small \frac{1}{4}\left(\begin{array}{cccc}
q_1&q_2&0&0\\
q_2&q_3&0&q_1\\
0&0&q_1&q_2\\
0&q_1&q_2&q_3\end{array}\right)},\ \  B_1=
{\small\frac{1}{4}\left(\begin{array}{cccc}
q_4&q_1&q_2&0\\
q_1&q_2&0&0\\
q_2&0&q_4&q_1\\
0&0&q_1&q_2\end{array}\right)}$$ and $$C_1=
{\small\frac{1}{4}\left(\begin{array}{cccc}
q_3&0&q_1&q_2\\
0&q_1&q_2&0\\
q_1&q_2&q_3&0\\
q_2&0&0&q_1\end{array}\right)},\ \  D_1=
{\small\frac{1}{4}\left(\begin{array}{cccc}
q_2&0&0&q_1\\
0&q_4&q_1&q_2\\
0&q_1&q_2&0\\
q_1&q_2&0&q_4\end{array}\right)}.$$ It is easy to check that
$A_1\geq0$ if and only if $q_1q_3\geq q_2^2$ and
$q_1^2q_3^2-2q_1q_2^2q_3-q_1^4+q_2^4\geq0$; $B_1\geq0$ if and only
if  $q_2q_4\geq q_1^2$ and
$q_2^2q_4^2-2q_1^2q_2q_4+q_1^4-q_2^4\geq0$; $C_1\geq0$ if and only
if  $q_1q_3^2\geq q_2^2q_3+q_1^3$ and
$q_1^2q_3^2-2q_1q_2^2q_3-q_1^4+q_2^4\geq0$; and $D_1\geq0$ if and
only if  $q_2q_4\geq q_1^2$ and
$q_2^2q_4^2-2q_1^2q_2q_4+q_1^4-q_2^4\geq0$.
Hence $$\begin{array}{rl} & \rho \ \mbox{\rm is PPT if and only if} \\
& q_1(q_1q_3^2-q_2^2q_3-q_1^3)\geq q_2^2(q_1q_3-q_2^2)\geq0 \\
\mbox{\rm and}&q_2(q_2q_4^2-q_1^2q_4-q_2^3)\geq
q_1^2(q_2q_4-q_1^2)\geq0.\end{array}\eqno(4.4)$$ Particularly,
$$\mbox{\rm if} \ q_2=2q_1 \ \mbox{and }\ q_3=q_4\geq 4q_1,\  \mbox{then} \ \rho \ \mbox{ is PPT
entangled. }\eqno(4.5) $$ This fact will be used below.

Now, let us apply  the realignment criterion to $\rho$. The
realignment
 of $\rho$ is
$$\begin{array}{rl}\rho^R\cong&\frac{1}{4}{\rm
diag}(q_1,q_1,q_1,q_1,q_1,q_1,q_1,q_1,q_1,q_1,q_1,q_1,q_1,q_1,q_1,q_1)\\
&+\frac{q_2}{4}(F_{1,16}+F_{2,13}+F_{3,14}+F_{4,15}+F_{5,4}+F_{6,1}+F_{7,2}+F_{8,3}\\
&+F_{9,8}+F_{10,5}+F_{11,6}+F_{12,7}+F_{13,12}+F_{14,9}+F_{15,10}+F_{16,11})\\
&+\frac{q_3}{4}(F_{1,11}+F_{6,16}+F_{11,1}+F_{16,6})
+\frac{q_4}{4}(F_{1,6}+F_{6,11}+F_{11,16}+F_{16,1})\\
\cong &A\oplus B^{(3)}\oplus 0,\end{array}$$ where

$$A=\frac{1}{4}\left(\begin{array}{cccc}
q_1&q_4&q_3&q_2\\
q_2&q_1&q_4&q_3\\
q_3&q_2&q_1&q_4\\
q_4&q_3&q_2&q_1\end{array}\right),\quad
B=\frac{1}{4}\left(\begin{array}{cccc}
q_1&0&0&q_2\\
q_2&q_1&0&0\\
0&q_2&q_1&0\\
0&0&q_2&q_1\end{array}\right)
$$
and $B^{(3)}$ denotes the direct sum of 3 copies of $B$. Then
$$\begin{array}{rl}\|\rho^R\|_1=&\|A\|_1+ 3\|B\|_1\\=&
\frac{3}{4}\sqrt{\sum_{i=1}^4q_i^2-q_1q_2-q_2q_3-q_3q_4-q_1q_4}\\
&+\frac{1}{4}\sqrt{\sum_{i=1}^4q_i^2+3(q_1q_2+q_2q_3+q_3q_4+q_1q_4)}\\
&+\frac{9}{4}\sqrt{q_1^2+q_2^2-q_1q_2}
+\frac{3}{4}\sqrt{q_1^2+q_2^2+3q_1q_2}.\end{array}\eqno(4.6)$$ Now a
computation reveals  that,  if $q_1\leq\frac{1}{15}$, $q_2=2q_1$ and
$q_3=q_4$, then the trace norm $\|\rho^R\|_1<1$. Note that, by
Eq.(4.5), $\rho$ is PPT in this case.  Hence,  we get another kind
of examples of entangled states that are PPT and cannot be detected
by the realignment criterion.

Similarly, by using the positive map  $\Phi^{\prime\prime}$, we
obtain that $\rho$ is entangled if $q_2<q_1$ or $q_3<q_2$, and, if
$q_2\leq\frac{1}{15}$, $q_1=2q_2$ and $q_3=q_4$, then $\rho$ is PPT
entangled that cannot be detected by the realignment criterion.

By using the positive map $\Phi^\prime$, we see that $\rho$ is
entangled if $q_3<q_1$ or $q_4<q_2$. In this case, by Eq.(4.4),
$\rho$ is not PPT because $q_1q_3^2-q_2^2q_3-q_1^3<0$ or
$q_2q_4^2-q_1^2q_4-q_2^3<0$.

\section{Positive finite rank elementary operators of order $(n,n)$}

In this section we consider the general case, that is, constructing
 positive finite rank elementary operators of  order $(n,n)$. The
main purpose is to show that the following result is true.

{\bf Theorem 5.1.} {\it Let $H$ and $K$ be  Hilbert spaces of
dimension $\geq n$, and let $\{|i\rangle\}_{i=1}^n$ and
$\{|j'\rangle\}_{j=1}^n$ be any orthonormal sets of $H$ and $K$,
respectively. For $k=1,2,\cdots, n-1$, let $\Phi^{(k)}:{\mathcal
B}(H)\rightarrow {\mathcal B}(K)$ be defined by
$$\begin{array}{rl} \Phi^{(k)}
(A)=&(n-1)\sum_{i=1}^nE_{ii}AE_{ii}^\dagger+\sum_{i=1}^{n}E_{i,\pi^k(i)}AE_{i,\pi^k(i)}^\dagger\\&-
(\sum_{i=1}^nE_{ii})A(\sum_{i=1}^nE_{ii})^\dagger
\end{array}\eqno(5.1)$$
for every $A\in{\mathcal B}(H)$, where $\pi(i)=\pi^1(i)=(i+1)\ {\rm
mod} \ n$, $\pi^k(i)=(i+k)\ {\rm mod} \ n$ ($k>1$), $i=1,2,\cdots,
n$ and $E_{ji}=|j'\rangle\langle i|$. Then $\Phi^{(k)}$ are positive
but not completely positive. Moreover, $\Phi^{(k)}$ is
indecomposable whenever either $n$ is odd or $k\not=\frac{n}{2}$.}

{\bf Proof.} Obviously, $\Phi^{(k)}$ is not completely positive for
each $k=1,2, \cdots ,n-1$. Similar to the proof of Theorem 4.1, to
prove that $\Phi=\Phi^{(1)}$ is positive, it is sufficient to show
that the function
$$\begin{array}{rl} &f_{1,n}(r_1,r_2,\cdots
,r_n)\\=&\left|\begin{array}{cccccccc} (n-2)r_1^2+r_2^2 & -r_1r_2 & -r_1r_3 &\cdots & -r_1r_n\\
-r_1r_2 & (n-2)r_2^2+r_3^2 & -r_2r_3 &\cdots & -r_2r_n\\
-r_1r_3 & -r_2r_3 & (n-2)r_3^2+r_4^2 &\cdots & -r_3r_n\\
\vdots&\vdots&\vdots&\ddots&\vdots\\
 -r_1r_n &-r_2r_n &-r_3r_n&\cdots&(n-2)r_n^2+r_1^2\end{array}\right|\geq 0\end{array} \eqno(5.2)$$
 for all $(r_1,r_2,\cdots ,r_n)$ with $0\leq
r_1,r_2,\cdots , r_n\leq 1$ and $\sum_{i=1}^nr_i^2=1$. Other
$\Phi^{(k)}$s are dealt with similarly.

We may assume that all $r_i$s are nonzero. Let
$x_i=\frac{r_{i+1}^2}{r_i^2}$, $i=1,2,\ldots , n-1$, and
$x_n=\frac{r_1^2}{r_n^2}$. Then $x_1x_2\cdots x_n=1$ and
$$f_{1,n}(r_1,r_2,\cdots ,r_n)=(r_1r_2\cdots r_n)^2
h_{1,n}(x_1,x_2,\cdots ,x_n),\eqno(5.3)$$ where
$$\begin{array}{rl} & h_{1,n}(x_1,x_2,\cdots
,x_n)\\=&\left|\begin{array}{cccccccc} (n-2)+x_1 & -1& -1 &\cdots & -1\\
-1 & (n-2)+x_2 & -1 &\cdots & -1\\
-1 & -1 & (n-2)+x_3 &\cdots & -1\\
\vdots&\vdots&\vdots&\ddots&\vdots\\
 -1 &-1 &-1&\cdots&(n-2)+x_n\end{array}\right| \end{array}\eqno(5.4)$$
with each $x_i>0$ and $x_1x_2\cdots x_n=1$. It follows that
$f_{1,n}\geq 0$ for all $(r_1,r_2,\cdots ,r_n)$ with $0\leq
r_1,r_2,\cdots , r_n\leq 1$ and $\sum_{i=1}^nr_i^2=1$ if and only if
$h_{1,n}\geq 0$ holds for all $(x_1,x_2,\cdots ,x_n)$ with $x_i>0$
($i=1,2,\ldots ,n$) and $x_1x_2\cdots x_n=1$.

Note that, the determinant in Eq.(5.4) can be formulated as
$$\begin{array}{rl}h_{1,n}(x_1,x_2,\cdots
,x_n)=&-M_0+M_1\sum_{i=1}^nx_i+M_2\sum_{i<j}x_ix_j+\cdots\\ &
+M_k\sum_{i_1<i_2<\cdots <i_k}x_{i_1}x_{i_2}\cdots x_{i_k} +\cdots
\\&
+M_{n-1}\sum_{i_1<i_2<\cdots <i_{n-1}}x_{i_1}x_{i_2}\cdots
x_{i_{n-1}}+M_nx_1x_2\cdots x_n. \end{array}$$

The case of $n=3$ is obvious. So we assume that $n\geq 4$ in the
sequel. Since, by Proposition 2.6, $h_{1,n}(0,0,\cdots, 0)=-M_0<0$,
we have $M_0>0$. By taking $x_i=0$ for $2\leq i\leq n$, it is easily
checked that $M_1=h_{1,n-1}(1,1,\cdots, 1)$. Let $x_i=0$ for $i\geq
3$. A computation reveals that $M_2=h_{1, n-2}(2,2,\cdots, 2)\geq
0$.  In general, one can check that
$$M_k=h_{1,n-k}(k,k,\cdots k)\geq 0, \quad k=1,2, \cdots,
n.\eqno(5.5)$$ For example,
$$M_{n-3}=h_{1,3}(n-3,n-3, n-3)=\left|\begin{array}{ccc}
n-2&-1&-1\\-1&n-2&-1\\-1&-1&n-2\end{array}\right|=(n-2)^3-3(n-2)-2\geq0,$$
$$M_{n-2}=h_{1,2}(n-2,n-2)=\left|\begin{array}{cc}
n-2&-1\\-1&n-2\end{array}\right|=(n-2)^2-1\geq 0,$$
$M_{n-1}=h_{1,n-1}=n-2\geq 0$ and $M_n=1$. Thus we have shown that
$M_0,M_1,M_2,\cdots M_n\in{\mathbb N}\cup\{0\}$. It is easily
checked that $h_{1,n}(1,1,\cdots, 1)=0$. This leads to
$$\sum_{i=1}^nM_i=M_0.\eqno(5.6)$$

Next, observe that if $a_j>0$ and $a_1a_2\cdots a_m=1$, then
$\sum_{j=1}^ma_j\geq 1$. It follows that
$$\sum_{i_1<i_2<\cdots <i_k}x_{i_1}x_{i_2}\cdots x_{i_k}\geq 1\eqno(5.7)$$ holds
for each $1\leq k\leq n$. Eq.(5.7), together with Eq.(5.6), yields
that $h_{1,n}(x_1,x_2,\cdots ,x_n)\geq 0$ holds for all
$(x_1,x_2,\cdots ,x_n)$ with $x_1x_2\cdots x_n=1$.

The last assertion will be proved by Example  5.4 below. The proof
is finished.\hfill$\Box$

{\bf Remark 5.2.} Let $\pi$ be any permutation of $(1,2,\cdots ,n)$
and let $\Psi_\pi :M_n({\mathbb C})\rightarrow M_n({\mathbb C})$ be
the map defined by
$$ \Psi_\pi (A)={\rm diag} \{(n-1)a_{11}+a_{\pi(1)\pi(1)},(n-1)a_{22}+a_{\pi(2)\pi(2)},\cdots ,(n-1)a_{nn}+a_{\pi(n)\pi(n)}\} -A
$$
for every $A=(a_{ij})\in M_n({\mathbb C})$. By Theorem 2.1,
Proposition 2.7 and the proof of Theorem 5.1,  it is  easily seen
that $\Psi_\pi$ is a positive linear map that is not completely
positive whenever $\pi\neq {\rm id}$.

{\bf Remark 5.3.} For any $n$-dimensional Hilbert space $H$, define
$$J_k=\frac{1}{\sqrt{k(k+1)}}(\sum_{i=1}^{k-1}E_{ii}-(k-1)E_{kk}),\ \ \ \ \ k=1,2,\cdots,n-1,$$
$$J_s=\left\{\begin{array}{cc} \frac{1}{\sqrt{2}}(E_{ij}+E_{ji}),\ \ {\rm if \ k \ is\ odd\ and \ i< j},\\
 \frac{1}{\sqrt{2}}(iE_{ij}-iE_{ji}),\ \ {\rm if \ k \ is \ even\ and \ i<j}.  \end{array}\right.$$
Relabel these $n^2-1$ matrices as $J_1,J_2,\cdots,J_{n^2-1}$. Then
the $n^2-1$ matrices form a completely orthonormal traceless set and
any $n\times n$ Hermitian matrix $S$ can be written as the form
$$S=\frac{1}{n}(I+ \sum_{k=1}^{n^2-1}\eta_kJ_k),$$
where $\eta_k\in{\mathbb R}$, $k=1,2,\cdots,n^2-1$. Hence it is
clear that the $n\times n$ hermitian matrices with trace 1 and the
points in ${\mathbb R}^{n^2-1}$ (the real linear space) are in
one-to-one correspondence. The image $\Lambda_n$ of the set of all
density matrices is a closed convex set in ${\mathbb R}^{n^2-1}$.
Then every positive linear map $\Phi:M_n({\mathbb C})\rightarrow
M_n({\mathbb C})$ corresponds to a linear map $M_{\Phi}:{\mathbb
R}^{n^2-1}\rightarrow{\mathbb R}^{n^2-1}$ that sends $\Lambda _n$
into $\Lambda_n$. It was shown in \cite{SRS} that every map
represented by a matrix of the form $M=(n-1)^{-1}R$ is positive,
where $R\in{\mathcal O}(n^2-1)$, the orthogonal group of proper and
improper rotations in ${\mathbb R}^{n^2-1}$ (\cite[Theorem 4]{SRS}).
Some more can be said. In fact, $M=(n-1)^{-1}R$ corresponds a
positive map whenever $\|R\|\leq 1$. The positive maps  in Theorem
3.1 may be obtained from this way. However, the positive maps  in
Theorem 4.1 can not be obtained from this way. For example, consider
the map $\Phi$ in Theorem 4.1. By a simple calculation, we get
$$\begin{array}{rl}M_{\Phi}=\frac{1}{18}\left(\begin{array}{ccc}
9&3\sqrt{3}&0\\
-\sqrt{3}&11&4\sqrt{2}\\
-2\sqrt{6}&-2\sqrt{2}&10\end{array}\right).
\end{array}$$
It is clear that $\|M\|>\frac{1}{3}$, and so \cite[Theorem 4]{SRS}
is not applicable to our map $\Phi$ here.

In the following we give two examples that generalize the examples
in Sections 3-4.

The states $\rho$ in Example 5.4 were suggested in \cite{JB} without
analyzing their entanglement.

{\bf Example 5.4.} Let $H$ and $K$ be Hilbert spaces of dimension
$\geq n$ and let $\{|i\rangle\}_{i=1}^n$ and
$\{|j^\prime\rangle\}_{j=1}^n$ be any orthonormal sets of $H$ and
$K$, respectively. Let
$|\omega\rangle=\frac{1}{n}\sum_{i=1}^n|ii'\rangle.$ Define
$\rho_1=|\omega\rangle\langle\omega|$,
$\rho_2=\frac{1}{n}\sum_{i=1}^{n}(I\otimes S)|ii'\rangle\langle
ii'|(I\otimes S)^\dagger,$
$\rho_3=\frac{1}{n}\sum_{i=1}^{n}(I\otimes S^2)|ii'\rangle\langle
ii'|(I\otimes S^2)^\dagger,$ $\ldots$,
$\rho_n=\frac{1}{n}\sum_{i=1}^{n}(I\otimes
S^{n-1})|ii'\rangle\langle ii'|(I\otimes S^{(n-1)})^\dagger$, where
$S$ is the operator on $K$ defined by $S|j'\rangle=|(j+1)'\rangle$
if $j=1,2,\cdots, n-1$, $S|n'\rangle=|1'\rangle$ and $S|j'\rangle=0$
if $j>n$. Let $\rho=\sum_{i=1}^nq_i\rho_i$ and
$\rho_t=(1-t)\rho+t\rho_0$, where
 $q_i\geq 0$ for $i=1,2,\cdots,n$ with $\sum_{i=1}^nq_i=1$, $t\in[0,1]$,
and $\rho_0$ is a state on $H\otimes K$. Then for sufficiently small
$t$, or for $\rho_0$ with $(\Phi^{(k)}\otimes I)\rho_0=0$
$k=1,2,\cdots ,n-1$, the following statements are true.

(1) If $q_i<q_1$ for some $i=2,3,\cdots,n$, then $\rho_t$ is
entangled;

(2) Let $\rho_0$ be PPT.  Then $\rho_t$ is a PPT state if and only
if $q_iq_j\geq q_1^2$ for $i,j$ with $i+j=n+2$, $i=3,4,\cdots,n$.

It is enough  to discuss   the entanglement of $\rho$. For
$\rho=\sum_{i=1}^nq_i\rho_i$, by using the map $\Phi=\Phi^{(1)}$ in
Theorem 5.1, it is easily checked that
$$\begin{array}{rl}&n(\Phi\otimes I)(\rho)\\
\cong&\left(\begin{array}{ccccc}
(n-2)q_1+q_n&-q_1&-q_1&\cdots&-q_1\\
-q_1&(n-2)q_1+q_n&-q_1&\cdots&-q_1\\
-q_1&-q_1&(n-2)q_1+q_n&\cdots&-q_1\\
\vdots&\vdots&\vdots&\ddots&\vdots\\
-q_1&-q_1&-q_1&\cdots&(n-2)q_1+q_n\end{array}\right)\\
&\oplus ( (n-2)q_n+q_{n-1})I_n \oplus
((n-2)q_{n-1}+q_{n-2})I_n\oplus\cdots\oplus ((n-2)q_2+q_1)I_n\oplus
0.
\end{array}$$
Thus, by Proposition 2.6, we get that $\rho$ is entangled if
$q_n<q_1$.

Similarly, by applying the map $\Phi^{(k)}$ in Theorem 5.1, we have
$\rho$ is entangled if $q_{n+1-k}<q_1$, where $k=2,3,\cdots, n-1$.

It is easily checked that $\rho$ is PPT if and only if $q_iq_j\geq
q_1^2$, where $i+j=n+2$ and $i=3,4,\cdots,n$.

Moreover, if $n$ is odd, or if $n$ is even but $k\not=\frac{n}{2}$,
we can choose $q_1,q_2,\cdots q_n$ so that
$q_{n+1-k}<q_1<\frac{1}{n}$ and $q_{i} q_{j}\geq q_i^2$ whenever
$i+j=n+2$. It follows that $\rho=\sum_{i=1}^nq_i\rho_i$ is PPT
entangled which can be recognized by $\Phi^{(k)}$. Hence,
$\Phi^{(k)}$ is not decomposable. This completes the proof of the
last assertion of Theorem 5.1.

{\bf Example 5.5.} Let $H$ and $K$ be complex Hilbert spaces of
dimension $\geq n$ and let $\{|i\rangle\}_{i=1}^{\dim H}$ and
$\{|j^\prime\rangle\}_{j=1}^{\dim K}$ be any orthonormal bases of
$H$ and $K$, respectively. Let
$|\omega_1\rangle=\frac{1}{\sqrt{n}}\sum_{i=1}^n|ii'\rangle$ and
$|\omega_{2}\rangle=\frac{1}{\sqrt{n}}(|12'\rangle+|23'\rangle+\cdots+|(n-1)n'\rangle+|n1'\rangle)$.
Define $\rho_1=|\omega_1\rangle\langle\omega_1|$,
$\rho_2=|\omega_2\rangle\langle\omega_2|$,
$\rho_3=\frac{1}{n}\sum_{i=1}^{n}(I \otimes S^2)|ii'\rangle\langle
ii'|(I \otimes S^{2\dagger}),$ $\ldots$,
$\rho_n=\frac{1}{n}\sum_{i=1}^{n}(I \otimes
S^{n-1})|ii'\rangle\langle ii'|(I \otimes S^{(n-1)\dagger})$, where
$S$ is the  same operator as in Example 5.4. Let
$\rho=\sum_{i=1}^nq_i\rho_i$ and $\rho_t=(1-t)\rho+t\rho_0$, where
$q_i\geq 0$ for $i=1,2,\cdots,n$ with $\sum_{i=1}^nq_i=1$,
$t\in[0,1]$, and $\rho_0$ is a state on $H\otimes K$. By using of
the positive finite rank elementary operators $\Phi^{(k)}$ in
Theorem 5.1, we can get that, for sufficient small $t$ or for any
$\rho_0$ with $(\Phi^{(k)}\otimes I)\rho_0=0$, $k=1,2,\cdots ,n-1$,
if $q_1\not=q_2$ or $q_1=q_2>q_i$ for some $i\in\{3,4,\cdots,n\}$,
then $\rho_t$ is entangled.

Still, we only need to consider the entanglement of $\rho$. For
$\rho=\sum_{i=1}^nq_i\rho_i$, with $\Phi=\Phi^{(1)}$ as in Theorem
5.1, it is clear that
$$\begin{array}{rl}n(\Phi\otimes I)(\rho)\cong &{\small \left(\begin{array}{ccccc}
(n-2)q_1+q_n&-q_1&-q_1&\cdots&-q_1\\
-q_1&(n-2)q_1+q_n&-q_1&\cdots&-q_1\\
-q_1&-q_1&(n-2)q_1+q_n&\cdots&-q_1\\
\vdots&\vdots&\vdots&\ddots&\vdots\\
-q_1&-q_1&-q_1&\cdots&(n-2)q_1+q_n\end{array}\right)}\\&\oplus
{\small\left(\begin{array}{ccccc}
(n-2)q_2+q_1&-q_2&-q_2&\cdots&-q_2\\
-q_2&(n-2)q_2+q_1&-q_2&\cdots&-q_2\\
-q_2&-q_2&(n-2)q_2+q_1&\cdots&-q_2\\
\vdots&\vdots&\vdots&\ddots&\vdots\\
-q_2&-q_2&-q_2&\cdots&(n-2)q_2+q_1\end{array}\right)}\\
&\bigoplus_{k=3}^n((n-2)q_k+q_{k-1})I_n\oplus 0\end{array}$$ So, by
Proposition 2.6,  $(\Phi\otimes I)(\rho)$ is not positive if
$q_n<q_1$ or $q_1<q_2$. It follows from the elementary operator
criterion that $\rho$ is entangled if $q_n<q_1$ or $q_1<q_2$.

Similarly, by applying the map $\Phi^{(k)}$ ($k=2,3,\cdots,n-1)$ in
Theorem 5.1, one gets that $\rho$ is entangled if $q_{n+1-k}<q_1$ or
$q_1<q_2$. Thus, we obtain that $\rho$ is entangled if $q_1\not=q_2$
or $q_1=q_2>q_i$ for some $i\in\{3,4,\cdots,n\}$.

Before the end of this section, we propose a question.

{\bf Question 5.6.} Let $n\geq 4$ be an even integer. Is the
positive map $\Phi^{(\frac{n}{2})}$ defined in Theorem 5.1
indecomposable? Particularly, is the positive map $\Phi^\prime$
defined in Theorem 4.1 indecomposable?

We guess that the answer is affirmative, but we are not able to
prove it here.

\section{Conclusions}

Let $H$ and $K$ be complex Hilbert spaces of any dimension. By the
elementary operator criterion \cite{H}, a state $\rho$ on $H\otimes
K$ is entangled if and only if there exists a positive finite rank
elementary operator $\Phi:{\mathcal B}(H)\rightarrow {\mathcal
B}(K)$ that is not completely positive (NCP) such that $(\Phi\otimes
I)\rho$ is not positive.  Hence it is important and interesting to
construct positive finite rank elementary operators that are NCP. In
this paper, we construct some new  positive finite rank elementary
operators and apply them to get some new examples of entangled
states. We also give a necessary and sufficient condition for a pure
state  to be separable in terms of a special positive elementary
operator of order $(2,2) $.

More concretely, for any positive integer $n\geq 3$, the  NCP
positive finite rank elementary operators that we constructed are
$\Phi^{(k)}:{\mathcal B}(H)\rightarrow {\mathcal B}(K)$ defined by
$\Phi^{(k)}
(A)=(n-1)\sum_{i=1}^nE_{ii}AE_{ii}^\dagger+\sum_{i=1}^{n}E_{i,\pi^k(i)}AE_{i,\pi^k(i)}^\dagger-
(\sum_{i=1}^nE_{ii})A(\sum_{i=1}^nE_{ii})^\dagger $ for every
$A\in{\mathcal B}(H)$, $k=1,2,\cdots, n-1$, where
$\{|i\rangle\}_{i=1}^n$ and $\{|j'\rangle\}_{j=1}^n$ are any
orthonormal sets of $H$ and $K$, respectively,
$E_{ji}=|j'\rangle\langle i|$ and $\pi^1=\pi$ is a permutation of
$\{1,2,\cdots ,n\}$ defined by $\pi(i)=(i+1)\ {\rm mod} \ n$,
$\pi^k(i)=(i+k)\ {\rm mod} \ n$ ($k>1$), $i=1,2,\cdots, n$.
Moreover, we show that $\Phi^{(k)}$ is indecomposable whenever
either $n$ is odd or $n$ is even but $k\neq \frac{n}{2}$. We discuss
two kinds of entangled states to illustrate how to use these
positive maps to detect the entanglement of states. Especially, we
study the examples in  detail for the case $n=4$ to determine when
they are PPT and when they can be detected by the realignment
criterion, and get some new examples of entangled states that cannot
be recognized by the PPT criterion and
 the realignment criterion.



\begin{thebibliography}{99}


\bibitem{BZ} I. Bengtsson, K. Zyczkowski,
Cambridge University Press, Cambridge, 2006.


\bibitem{Hor} M. Horodecki, P. Horodecki, R. Horodecki,
Phys. Lett. A 223 (1996) 1.

\bibitem{Pe} A. Peres, 
Phys. Lett. A 202 (1996) 16.



\bibitem{CW} K. Chen, L. Wu, Quant. Inf. Comput 3 (2003) 193.

\bibitem{B} D. Bru${\ss}$, J. Math. Phys. 43  (2002) 4237.

\bibitem{TG} G. T$\acute{o}$th, O. G$\ddot{u}$hne, Phys. Rev. Lett. 94 (2005) 060501.

\bibitem{CK}  D. Chru$\acute{s}$ci$\acute{n}$ski and A. Kossakowski,
Open Systems and Inf. Dynamics 14 (2007) 275;  D.
Chru$\acute{s}$ci$\acute{n}$ski and A. Kossakowski, J. Phys. A:
Math. Theor. 41 (2008) 145301.

\bibitem{JB} M. A. Jafarizadeh, N. Behzadi, Y. Akbari,
Eur. Phys. J. D  55 (2009) 197.

\bibitem{HHH1} R. Horodecki, P. Horodecki, M. Horodecki,
Rev. Mod. Phys. 81 (2009) 865.




\bibitem{H5} J. Hou,
Sci. in China (ser.A), 36(9) (1993), 1025-1035.

\bibitem{H4} J. Hou,
J. Operator Theory, 39 (1998), 43-58.

\bibitem{HQ}  J. Hou, X. Qi,
Phys. Rev. A 81 (2010) 062351.

\bibitem{H} J. Hou,
J. Phys. A: Math. Theor. 43 (2010) 385201; arXiv[quant-ph]:
1007.0560v1.

\bibitem{NC} M. A. Nielsen, I. L. Chuang,
Cambridge University Press, Cambridge, 2000.

\bibitem{SRS} S. Simon, S. P. Rajagopalan, R. Simon,
Pramana-Journal of Physics, 73(3) (2009) 471-483.

 \bibitem{W} R. F. Werner,
Phys. Rev. A 40 (1989) 4277.





\end{thebibliography}
\end{document}